\documentclass[journal]{IEEEtran}

\ifCLASSINFOpdf
\else
\fi

\usepackage{cite}
\usepackage{mathrsfs}
\usepackage{amssymb}

\usepackage{amsmath}

\usepackage{hyperref}  
\usepackage[nameinlink,capitalise]{cleveref}

\usepackage{graphicx}
\usepackage{multirow} 
\usepackage{booktabs}
\usepackage[table]{xcolor}

\usepackage{array}
\usepackage[linesnumbered,ruled,vlined]{algorithm2e}
\let\oldnl\nl
\newcommand{\nonl}{\renewcommand{\nl}{\let\nl\oldnl}}

\usepackage[caption=false,font=normalsize,labelfont=sf,textfont=sf]{subfig}

\usepackage{caption}
\makeatletter
\newcommand{\thickhline}{%
    \noalign {\ifnum 0=`}\fi \hrule height 1pt
    \futurelet \reserved@a \@xhline
}
\newcolumntype{"}{@{\hskip\tabcolsep\vrule width 1pt\hskip\tabcolsep}}
\makeatother

\usepackage{nomencl}
\makenomenclature

\begin{document}


\title{Security-Constrained Substation Reconfiguration Considering Busbar and Coupler Contingencies}

%
%
%

\author{Ali~Rajaei,~\IEEEmembership{Graduate Student Member,~IEEE}, and~Jochen~L. Cremer,~\IEEEmembership{Senior Member,~IEEE}
\thanks{This work is supported by the TU Delft AI Labs \& Talent Programme. A. Rajaei, and J. Cremer are with the Department of Electrical Sustainable Energy, Delft University of Technology, Delft, The Netherlands, and also with the Center for Energy, Austrian Institute of Technology, Austria (e-mail: a.rajaei, j.l.cremer\{@tudelft.nl\}).}
}

\maketitle
\markboth{IEEE Transactions on Power Systems}{}

\begin{abstract}
Substation reconfiguration via busbar splitting can mitigate transmission grid congestion and reduce operational costs. However, existing approaches neglect the security of substation topology, particularly for substations without busbar splitting (i.e., closed couplers), which can lead to severe consequences. Additionally, the computational complexity of optimizing substation topology remains a challenge. This paper introduces a MILP formulation for security-constrained substation reconfiguration (SC-SR), considering N-1 line, coupler and busbar contingencies to ensure secure substation topology. To efficiently solve this problem, we propose a heuristic approach with multiple master problems (HMMP). A central master problem optimizes dispatch, while independent substation master problems determine individual substation topologies in parallel. Linear AC power flow equations ensure PF accuracy, while feasibility and optimality sub-problems evaluate contingency cases. The proposed HMMP significantly reduces computational complexity and enables scalability to large-scale power systems. Case studies on the IEEE 14-bus, 118-bus, and PEGASE 1354-bus system show the effectiveness of the approach in mitigating the impact of coupler and busbar tripping, balancing system security and cost, and computational efficiency.    
\end{abstract}


%
\IEEEpeerreviewmaketitle

\section*{Nomenclature}

\addcontentsline{toc}{section}{Nomenclature}

\begin{IEEEdescription}[\IEEEusemathlabelsep\IEEEsetlabelwidth{$P_abcd,Q_abd$}]
\item[Indices and Sets]
\item[$i \in \mathcal{V} $] Index for substations (nodes).
\item[$b \in \mathcal{B}$] Index for busbars $\mathcal{B}=\{b1,b2\}$.
\item[$ij \in \mathcal{L} $] Index for transmission lines (from $i$ to $j$).
\item[$g \in \mathcal{G} $] Index for generators.
\item[$d \in \mathcal{D} $] Index for load demands.
\item[$c \in \Omega^*$] Index for contingency states. $\Omega^l$: line contingencies, $\Omega^c$: coupler contingencies, $\Omega^b$: busbar contingencies. $c=0$ is normal state.
\item[$c \in \Omega_i$] Index for contingencies of elements of substation $i$, including the coupler $c_i$ and the busbars $b1_i,b2_i$.  
\item[$\mathcal{L}_i,\mathcal{G}_i,\mathcal{D}_i$] Elements connected to substation $i$.
\item[$\Omega^{cnd},\Omega^{split}$] Candidate and selected substations for splitting.

\item[Parameters]
\item[$\hat{P}_d,\hat{Q}_d$] Active and reactive power demand.
\item[$\underline{P}_g,\overline{P}_g$] Minimum and maximum active power generation.
\item[$\underline{Q}_g,\overline{Q}_g$] Minimum and maximum reactive power generation.
\item[$\overline{r}^u_g,\overline{r}^d_g$] Maximum ramp-up and ramp-down.
\item[$\underline{V}_i,\overline{V}_i$] Minimum and maximum squared voltage magnitude.
\item[$\overline{Z^s}$] Maximum number of busbar splitting. 
\item[$\hat{P}^{m}_g$] Active power generation cleared in the electricity market.
\item[$b_{ij},g_{ij}$] Susceptance and conductance of line $ij$. 
\item[$\overline{S}_{ij}$] Maximum apparent power flow of line $ij$.
\item[$\overline{S}_{i}$] Maximum apparent power flow of the coupler in substation $i$.
\item[$\phi_d$] Tangent of the power factor angle of loads. 
\item[$\pi_g,\pi^u_g, \pi^d_g,\pi_d$] Cost (€/MWh) of generation, ramp-up, ramp-down, and load shedding. 
\item[$M^\theta,M^v$] Big-M parameters.
\item[$\hat{x},\hat{z}$] Fixed dispatch and topology solution. 
\item[$\epsilon^f,\epsilon^o$] Feasibility and optimality thresholds.  
\item[$p_c$] Probability of contingency $c$.
\item[$\alpha$] Relative additional generation cost.

\item[Variables] 
\item[$P_{g,b,c},Q_{g,b,c}$] Active and reactive power generation for contingency states.
\item[$P^0_{g}$] Active power dispatch for normal state.
\item[$r^u_g,r^d_g$] Ramp-up and ramp-down spinning reserve.
\item[$P_{d,b},Q_{d,b}$] Active and reactive load demand.
\item[$P'_{d,b,c},Q'_{d,b,c}$] Active and reactive load shedding.
\item[$P_{i,c},Q_{i,c}$] Active and reactive power flow through the coupler at substation $i$ from busbar $b1$ to $b2$. 
\item[$P_{ij,c},Q_{ij,c}$] Active and reactive power flow of line $ij$ from $i$ side. 
\item[$P_{ij,b,c},Q_{ij,b,c}$] Active and reactive power flow of line $ij$ through busbar $b$ at substation $i$. 
\item[$P^L_{ij,c},Q^L_{ij,c}$] Active and reactive power loss of line $ij$.
\item[$\theta_{i,b,c},V_{i,b,c}$] Voltage angle and squared voltage magnitude of substation $i$ busbar $b$.
\item[$\theta_{ij,c},V_{ij,c}$] Voltage angle and squared voltage magnitude at the end of line $ij$ on the side of $i$. 
\item[$z_i$] Connection status of coupler in substation $i$ (0: open, 1: close). 
\item[$z_g, z_d$] Connection status of generator and demand to busbars ($0:b1, 1:b2$). 
\item[$z_{ij}$] Connection status of line $ij$ to substation $i$ busbars ($0:b1, 1:b2$).
\item[$\mu_{.,c}, \lambda_{.,c}$] Dual variables of Benders constraints for sub-problem $c$.
\item[$\mathcal{O}^{MP}_0,\mathcal{O}^{MP}_i$] Objective of the master problem for dispatch and topology of substation $i$.
\item[$\mathcal{O}^{FSP}_c,\mathcal{O}^{OSP}_c$] Objective of the feasibility and optimality sub-problems.
\item[$\psi_c$] Positive auxiliary variables for Benders cuts. 
\item[$UB,LB$] Upper bound and lower bound.

\end{IEEEdescription}

\normalsize

\section{Introduction}
%
%
%
%

\begin{figure}[t]
    \centering \includegraphics[width=0.38\textwidth, keepaspectratio=true,trim={0cm 8.5cm 18cm 0cm},clip]{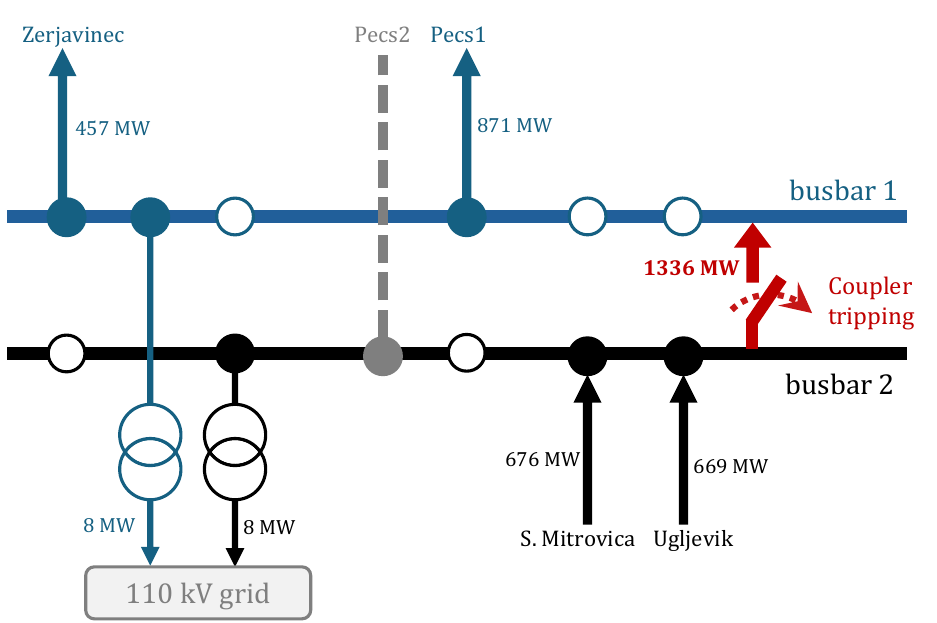}
    \caption{Topology in substation Ernestinovo prior to the system split on Jan-2021. The topology was not changed after the planned outage of line Pecs2. Switching the line Pecs1 to busbar 2 would have relieved the coupler and prevented the cascading failures \cite{panel2021continental}.}
    \label{fig:Ern_sub}
\end{figure}

The increasing share of distributed energy resources and the rapid electrification of various sectors pose challenges for system operators in alleviating the increased congestion in the grid. Transmission network topology control, particularly substation reconfiguration via busbar splitting, is a promising solution to alleviate congestion and reduce operational costs, offering an alternative to costly redispatch measures \cite{mazi1986corrective,shao2005corrective}. However, the security of substation topology is often overlooked, especially when busbar splitting is not applied. System operators rely on experience or predefined manuals for substation topology, while outages of substation elements are neglected in the contingency list, which can lead to severe consequences. On January 8, 2021, the outage of a single highly loaded busbar coupler (or ‘coupler’) triggered cascading failures that ultimately split the European grid into two areas \cite{panel2021continental}. This happened because the substation topology was not adjusted after a transmission line was taken out of service, and the N-1 coupler outage was not considered in the contingency analysis, as \cref{fig:Ern_sub} shows. To prevent such events, ENTSO-E recommends selecting substation topologies that minimize the flow through couplers and including all transmission elements, such as couplers, in the contingency list \cite{panel2021continental}. In this context, this paper proposes a computationally efficient and scalable approach to determine secure substation topologies considering busbars, couplers, and line contingencies, that can be integrated with current operational practices in large-scale grids, and probabilistic approaches or market applications to balance system security and operational cost.

Optimal transmission line switching (OTS), first formulated by Fisher et al. \cite{fisher2008OTS} as a MILP, extends the DC optimal power flow (OPF) by adding binary variables for line switching to reduce costs. As the OTS problem \cite{fisher2008OTS} is NP-hard \cite{lehmann2014complexity}, heuristic approaches, such as sensitivity analysis \cite{fuller2012fast,ruiz2012tractable} and constraint identification \cite{crozier2022feasible} are proposed. To address N-1 line contingencies, \cite{khanabadi2012OTS} uses Benders decomposition (BD) to handle violations, while \cite{ruiz2016SC-OTS} uses sensitivity factors to formulate a security-constrained (SC) OTS. Moreover, corrective line switching actions are explored in \cite{li2016real,li2019enhanced}. However, these works do not consider substation reconfiguration actions.

Optimal substation reconfiguration (OSR) involves adjusting substation topology, including busbar splitting and the reconfiguration of transmission feeders to the busbars (referred to as substation configuration) \cite{mazi1986corrective,shao2005corrective}. Several works, such as \cite{shao2005corrective,zhou2019bus,sogol2021tractable,sogol2023congestion, wang2023bus,van2023bus,bastianel2025identifying} use sensitivity analysis as a heuristic approach to identify candidate busbar splitting actions. These heuristic approaches assume a fixed generation dispatch and substation configuration, focusing solely on identifying busbar splitting actions to marginally improve their objective, but they do not provide an optimal substation configuration. \cite{heidarifar2015network} proposes a MILP formulation of OSR based on node-breaker modeling and DC OPF. Alternatively, \cite{goldis2016shift} uses shift factors to develop a MILP that incorporates substation reconfiguration in the SC-OPF problem. \cite{morsy2025configure} develops a heuristic search approach based on the locality effects of switching actions to solve the MIP. \cite{ewerszumrode2024iterative} proposes an iterative approach that decouples topology and redispatch optimization. \cite{heidarifar2021optimal} extends the work in \cite{heidarifar2015network} by incorporating N-1 line contingencies using BD. \cite{morsy2022security} advances these works by developing an SC-OPF with corrective busbar splitting actions using the column-and-constraint generation technique. \cite{bastianel2024ac-dc,bastianel2025day-ahead-ac-dc,morsy2025ac-dc} focus on hybrid AC/DC grids. In particular, \cite{morsy2025ac-dc} extends \cite{morsy2022security} by introducing soft busbar splitting in converter substations, while \cite{bastianel2025day-ahead-ac-dc} extends \cite{bastianel2024ac-dc} to day-ahead optimization considering renewable uncertainty. In addition, \cite{li2019day-ahead-UC} builds on \cite{heidarifar2015network} to consider unit commitment for day-ahead scheduling with dynamic thermal rating. 

A limitation of \cite{heidarifar2015network,goldis2016shift,morsy2022security,morsy2025configure,heidarifar2021optimal,bastianel2024ac-dc,bastianel2025day-ahead-ac-dc,morsy2025ac-dc,li2019day-ahead-UC,ewerszumrode2024iterative} is that they neglect the substation configuration when the coupler is closed (i.e., not split), effectively modeling the substation as a single electric node (e.g., forcing all elements to the same busbar). While this simplification does not affect the cost objective under normal or line contingency cases, it prevents a correct representation of coupler and busbar outages, whose impact depends on the allocation of elements across busbars. As a result, a topology that seems feasible under normal and line contingency cases may still be unsafe under coupler and busbar contingencies \cite{panel2021continental,ciapessoni2016probabilistic}. Moreover, the aforementioned works \cite{shao2005corrective,zhou2019bus,sogol2021tractable,sogol2023congestion, wang2023bus,van2023bus,bastianel2025identifying,heidarifar2015network,goldis2016shift,morsy2022security,morsy2025configure,heidarifar2021optimal,bastianel2024ac-dc,bastianel2025day-ahead-ac-dc,morsy2025ac-dc,li2019day-ahead-UC,ewerszumrode2024iterative} rely on DC PF equations, which can lead to switching actions that violate AC feasibility or inadvertently increase operational costs \cite{soroush2013accuracies}.

The AC OTS is explored in \cite{soroush2013accuracies,khanabadi2012OTS,capitanescu2014ac,poyrazoglu2015ACOTS,bai2016ACOTS,kocuk2017ACOTS,brown2020ACOTS-LP}. \cite{soroush2013accuracies} extends the DCOPF-based pre-screening heuristics from \cite{fuller2012fast,ruiz2012tractable} to ACOPF. In \cite{khanabadi2012OTS}, a BD approach with an AC feasibility sub-problem is presented. \cite{capitanescu2014ac} develops a heuristic approach that solves a sequence of AC OPF problems to identify one line removal at each step. Several AC OTS formulations using semidefinite programming relaxation \cite{poyrazoglu2015ACOTS}, second-order cone programming relaxation \cite{bai2016ACOTS,kocuk2017ACOTS}, and linear programming approximation \cite{brown2020ACOTS-LP} have been developed. Regarding AC OSR, \cite{park2018sparse} develops a sparse tableau AC PF formulation, and the resulting problem is solved using an outer approximation method in \cite{park2020optimal} and an iterative heuristic in \cite{sogol2025sparse}. \cite{heidarifar2021optimal} builds on \cite{heidarifar2015network} using BD, where a MISOCP master problem determines the topology and sub-problems ensure N-1 and AC feasibility. However, none of these works consider the contingencies of substation elements, such as couplers, which can lead to severe operational challenges, as observed in the European grid split in 2021 \cite{panel2021continental}.

This paper proposes a security-constrained substation reconfiguration (SC-SR) optimization that considers N-1 line, coupler and busbar contingencies. We use linear AC PF equations to ensure AC feasibility. To handle the computational complexity of SC-SR, we propose a heuristic approach with multiple master problems (HMMP). The approach decomposes the problem into a central master problem ($\mathit{MP}_0$) that determines generator dispatch, and independent substation master problems ($\mathit{MP}_i$) that optimize the topology of each substation in parallel. Feasibility and optimality sub-problems assess contingency cases and iteratively add cuts. 

Existing decomposition approaches do not perform well for SC-SR as they process all binary topology decisions together, which leads to slow or stalled convergence due to symmetries in the problem. Sensitivity-based and sequential heuristics are developed for normal-state DC-PF problems, also process all binaries together, and depend on the sequence order, which prevents effective parallelization. HMMP is inspired by ideas used in BD, sequential switching heuristics, and problem partitioning, but the unified scheme (with multiple MPs and feasibility and optimality checks in an iterative scheme) is new and tailored to the SC-SR problem.

The main contributions of this paper are: \begin{itemize}
    \item The MILP formulation of SC-SR considering N-1 coupler and busbar contingencies, as well as linear AC PF, to determine a secure configuration for all substations. 
    \item The HMMP approach that achieves close-to-optimal solutions, substantially reduces the computational complexity, and facilitates scalability to large-scale power systems. 
    \item The HMMP approach to identify busbar splitting actions based on reducing objective value.
    \item Exploring use cases, including: (i) probabilistic security and (ii) fixed-cost approaches to balance system security and operational cost, and (iii) day-ahead scheduling of substation reconfiguration. 
\end{itemize}

The paper is organized as follows. \cref{sec2:formulation} introduces the proposed MILP formulation for SC-SR. \cref{sec3:solution} presents the HMMP solution approach and BD baselines. \cref{sec:results} presents the case studies, and \cref{sec:conc} concludes the paper.

\section{Security-Constrained Substation Reconfiguration (SC-SR) Problem}
\label{sec2:formulation}

\subsection{Problem Statement}
The general formulation of SC-SR can be stated as: 
\begin{equation}
    \begin{aligned}    
     \label{eq:problem-state}         
     \min \quad & f(x) + \sum_{c \in \Omega^*} g_c(y_c) 
     \\  \text{s.t} \quad & h_c(x,z,y_c,w_c) \leq 0 \quad c \in (\Omega^* +\{0\})  
     \end{aligned}
\end{equation}
where the preventive decision variables are the generation dispatch $x\in \mathbb{R}^{m_a}$, and the topology $ z \in \{0,1\}^{m_b}$. For each contingency state $c \in \Omega^*$ (and the normal state $c=0$), $w_c \in \mathbb{R}^{m_{c}}$ denotes operational variables (e.g., flows, voltages), and $y_c \in \mathbb{R}^{m_{d}}$ denotes load curtailments ($y_0=0$). The function $h(.)$ presents all AC PF and operational constraints, while $f(.)$ and $g_c (.)$ present generation cost and load shedding cost, respectively. The objective function minimizes total cost by optimizing dispatch and topology under a specified cost–security trade-off. The general formulation in \eqref{eq:problem-state} can be adapted to different cost–security paradigms. In the case studies, we explore two examples: a probabilistic security objective and a fixed-cost approach.

\subsection{Substation Model}
Transmission network substations act as junction points for system elements, including transmission lines, distribution feeders, and generators. Adjusting the configuration of busbars and controlling circuit breakers (CBs) enables changes in substation configuration, which alter the connection between system elements and the overall network topology. This paper focuses on two commonly used substation arrangements: the double-bus double-breaker and the breaker-and-a-half arrangements \cite{isoNE2017,Book_sub_layout, heidarifar2015network,heidarifar2021optimal}. We extend the model in \cite{heidarifar2015network} to consider coupler and busbar contingencies, as well as linear AC PF constraints.

\begin{figure}[t]
    \centering \includegraphics[width=0.42\textwidth, keepaspectratio=true,trim={0cm 10.0cm 17.9cm 0cm},clip]{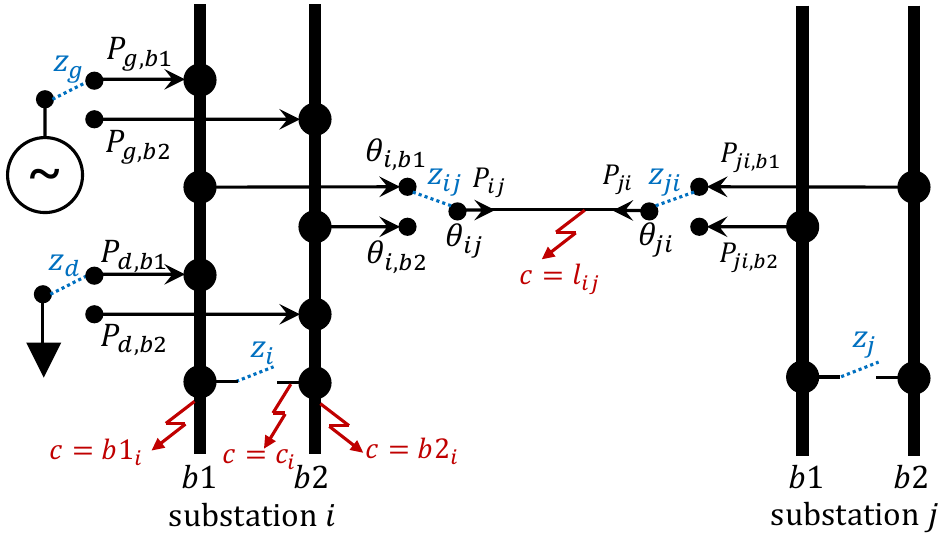}
    \caption{Schematic of the generalized substation model and different type of contingencies.}
    \label{fig:substation}
\end{figure}

\cref{fig:substation} shows the generalized substation model and different type of contingencies. Substation $i$ is connected to substation $j$ through a transmission line $l_{ij}$. Elements connected to each substation, e.g. lines, can be switched to either of the busbars $b1$ or $b2$. To this end, binary variables $z_{ij}, z_g, z_d=0$ represent connection to $b1$, and $z_{ij}, z_g, z_d=1$ represent connection to $b2$. The binary variable $z_i$ indicates the coupler status, where $z_i=0$ represents splitting. Additionally, three types of contingencies are shown: line contingencies $c=l_{ij} \in \Omega^l$, coupler contingencies $c=c_i \in \Omega^c$, and busbar contingencies $c=b1_i,b2_i \in \Omega^b$. A coupler outage results in unintended busbar splitting, whereas a busbar outage leads to simultaneous outages of all elements connected to that busbar \cite{ciapessoni2016probabilistic,Book_sub_layout}.

\subsection{The Proposed MILP Formulation}
The proposed MIP formulation for substation reconfiguration considers busbar and coupler contingencies, and linear AC PF equations. 

\subsubsection{Substations}
The voltage and the coupler flow constraints are as below, which apply to $i \in \mathcal{V}, c \in \Omega^*$: 

\begin{subequations}
\label{eqCO-sub}
\begin{align}    
& -M^\theta(1-z_i) \leq \theta_{i,b1,c}-\theta_{i,b2,c} \leq M^\theta(1-z_i)  \quad  c\notin \Omega_i  \label{eq:sub-theta} \\ 
& -M^v(1-z_i) \leq V_{i,b1,c}-V_{i,b2,c} \leq M^v(1-z_i)  \quad  c \notin \Omega_i \label{eq:sub-v} \\
&  P^2_{i,c}+Q^2_{i,c} \leq z_i \overline{S}^2_i \quad  c\notin \Omega_i \label{eq:sub-linelimit} \\
    & P_{i,c}=0,\quad Q_{i,c}=0 \quad c \in \Omega_i \label{eq:sub-0}   
\end{align}
\end{subequations}
where $\Omega^*$ denotes all contingency states, and $\Omega_i=\{c_i,b1_i,b2_i\}$ are contingencies of the substation's elements, i.e., the coupler $c_i$ and the busbars $b1_i,b2_i$. \eqref{eq:sub-theta}-\eqref{eq:sub-v} enforce equal voltage angle and magnitude when the coupler is closed and no contingency has occurred on the substation ($c \notin \Omega_i$). \eqref{eq:sub-linelimit} is the thermal limit of the coupler flow. In the event of a contingency $c \in \Omega_i$, the coupler flow is set to zero.

The transmission lines connected to each substation $i$ can be assigned to either of the busbars through the binary variable $z_{ij}$, which is modeled as follows for $l_{ij} \in \mathcal{L}_i, c \in \Omega^*$:  
\begin{subequations}
\label{eqCO-lines}
\begin{align}
& -z_{ij}M^\theta \leq \theta_{ij,c} - \theta_{i,b1,c} \leq z_{ij}M^\theta \quad  c\neq l_{ij}, c \neq b1_i  \label{eq:lines-theta0}\\   
& -z_{ij}M^v \leq V_{ij,c} - V_{i,b1,c} \leq z_{ij}M^v \quad  c\neq l_{ij}, c \neq b1_i \label{eq:lines-v0} \\
& -(1-z_{ij})\overline{S}_{ij} \leq P_{ij,b1,c}  \leq (1-z_{ij})\overline{S}_{ij}  \label{eq:lines-P0} \\
& -(1-z_{ij})\overline{S}_{ij} \leq Q_{ij,b1,c}  \leq (1-z_{ij})\overline{S}_{ij} \label{eq:lines-Q0} 
\\
& -(1-z_{ij})M^\theta \leq \theta_{ij,c} - \theta_{i,b2,c} \leq (1-z_{ij})M^\theta \nonumber \\ &\quad  c\neq l_{ij}, c \neq b2_i  \label{eq:lines-theta1} \\
& -(1-z_{ij})M^v \leq V_{ij,c} - V_{i,b2,c} \leq (1-z_{ij})M^v \nonumber \\& \quad  c\neq l_{ij}, c \neq b2_i  \label{eq:lines-v1} \\
& -z_{ij}\overline{S}_{ij} \leq P_{ij,b2,c}  \leq z_{ij}\overline{S}_{ij} \label{eq:lines-P1} \\
& -z_{ij}\overline{S}_{ij} \leq Q_{ij,b2,c}  \leq z_{ij}\overline{S}_{ij} \label{eq:lines-Q1} 
\\
& P_{ij,c}=P_{ij,b1,c}+P_{ij,b2,c} \\
& Q_{ij,c}=Q_{ij,b1,c}+Q_{ij,b2,c} \\
& P_{ij,b,c}=0,\quad Q_{ij,b,c}=0 \quad b \in \mathcal{B}, (c=b_i \vee c=l_{ij}) \label{eq:lines-PQcont}
\end{align}
\end{subequations}
where $\theta_{i,b}$ is the voltage angle of the substation on each busbar $b$, and $\theta_{ij}$ is the voltage angle of line $ij$ on the $i$ side, as \cref{fig:substation} shows. Note that $\theta_{ij}$ is not the angle difference of the line $ij$. As an example, when $z_{ij}=0$, \eqref{eq:lines-theta0}-\eqref{eq:lines-v0} and \eqref{eq:lines-P1}-\eqref{eq:lines-Q1} are binding to enforce the connection to busbar $b1$ and zero flow to $b2$, unless a busbar or line outage occurs ($c=b_i \vee c=l_{ij}$), in which case \eqref{eq:lines-PQcont} sets the flows to zero. 

Constraints \eqref{eq:zmax}-\eqref{eq:z-sym2} consider operational assumptions and improve the computational complexity:
\begin{subequations}
\label{eqCO-sw}
    \begin{align}
        &\sum_{i \in \mathcal{V}} (1-z_i) \leq \overline{Z^s} \label{eq:zmax}\\
        &2(1-z_i) \leq \sum_{ij \in \mathcal{L}_i}z_{ij} \quad i \in \mathcal{V} 
\label{eq:z-sym0} \\
        &2(1-z_i) \leq \sum_{ij \in \mathcal{L}_i}(1-z_{ij}) \quad i \in \mathcal{V} \label{eq:z-sym1} \\
        & z_{ij}=0 \quad i \in \mathcal{V}, l_{ij}=\text{argmin}(\mathcal{L}_i) \label{eq:z-sym2}
    \end{align}
\end{subequations}
where \eqref{eq:zmax} limits the total number of busbar splitting actions, as set by the system operator. \eqref{eq:z-sym0}-\eqref{eq:z-sym1} ensure system security by enforcing that at least two lines are connected to each busbar if that substation splits, which prevents busbar isolation due to a line contingency \cite{heidarifar2015network, heidarifar2021optimal}. Furthermore, we introduce \eqref{eq:z-sym2} to eliminate the symmetry in the formulation by fixing one line (here the line with the lowest index) at each substation to $b1$. This constraint reduces the number of possible combinations by half (see Appendix).   

In the event of a coupler or busbar outage, substation configuration significantly impacts resulting load loss. Thus, this paper determines the secure configuration of all substations, not only those that are split. In contrast, previous works such as \cite{heidarifar2015network,heidarifar2021optimal,morsy2022security} apply bounds (e.g., $z_i + z_g \leq 1$) that force all elements to one busbar when the substation is not split, assuming it acts as a single node. This assumption is valid under normal or line contingency conditions and is used to remove the symmetry in the formulation \cite{heidarifar2015network}. However, as substation contingencies are considered through \eqref{eqCO-sub}-\eqref{eqCO-lines}, the busbar assignment of all elements affects the post-contingency state, and this symmetry no longer exists.

\subsubsection{Generators and loads} 
Constraints for generators are: 
\begin{subequations}
\label{eqCO-gen}
\begin{align}
    &(1-z_g)\underline{P}_g \leq P_{g,b1,c} \leq (1-z_g)\overline{P}_g &g \in \mathcal{G}, c \in \Omega^*  \label{eq:g0}\\
    &z_g\underline{P}_g \leq P_{g,b2,c} \leq z_g\overline{P}_g  &g \in \mathcal{G}, c \in \Omega^* \label{eq:g1} \\
    &(1-z_g)\underline{Q}_g \leq Q_{g,b1,c} \leq (1-z_g)\overline{Q}_g  &g \in \mathcal{G}, c \in \Omega^* \label{eq:g2}\\
    &z_g\underline{Q}_g \leq Q_{g,b2,c} \leq z_g\overline{Q}_g &g \in \mathcal{G}, c \in \Omega^* \label{eq:g3} \\
    & P^0_{g,b} - r^d_g \leq P_{g,b,c} \leq P^0_{g,b} + r^u_g &g \in \mathcal{G}, c \in \Omega^* \label{eq:g5}\\
    & P^0_{g}=P^0_{g,b1}+P^0_{g,b2}  &g \in \mathcal{G} \\
    & 0 \leq r^u_g \leq \overline{r}^u_g, \quad 0 \leq r^d_g \leq \overline{r}^d_g  &g \in \mathcal{G} \label{eq:g6}
\end{align}
\end{subequations}
where \eqref{eq:g0}-\eqref{eq:g3} determine the connection status of generators by $z_g$ to either of the substation’s busbars. \eqref{eq:g5} limits the generator output under contingency states based on the purchased spinning reserves, and \eqref{eq:g6} constrains the spinning reserves based on the generators' ramping requirements \cite{davoodi2022methodology}.     

Constraints for load demands are:
\begin{subequations}
\label{eqCO-dem}
\begin{align}
    &P_{d,b1} = (1-z_d)\hat{P}_d  
    ,\quad P_{d,b2}=z_d \hat{P}_d  \quad d \in \mathcal{D} \label{eq:d0}\\
    &P'_{d,b,c} \leq P_{d,b} \quad c \in \Omega^* \quad d \in \mathcal{D} \label{eq:d1}\\
    &Q_{d,b}=\phi_d P_{d,b}, \quad  Q'_{d,b,c}=\phi_dP'_{d,b} \quad d \in \mathcal{D}, c \in \Omega^* \label{eq:d2} 
\end{align}
\end{subequations}
where \eqref{eq:d0} determines the connection status of loads by $z_d$. \eqref{eq:d1} limits the load shedding by the actual demand, while \eqref{eq:d2} defines reactive load (shedding) assuming constant power factor \cite{alizadeh2022tractable}. 

\subsubsection{Linear AC power flow} 
AC PF equations are essential for accurately modeling power flow in complex systems, while DC PF equations overlook reactive power and voltage constraints, leading to infeasible or sub-optimal solutions. However, using full AC equations results in a computationally expensive MINLP problem. In this paper, we adopt the linear approximation of AC equations in \cite{yang2017linearized,alizadeh2022tractable}, achieving a balance of accuracy and efficiency in a MILP formulation. The linear AC PF constraints apply to $c \in \Omega^*$:
\begin{subequations}
\label{eqCO-PF}
\begin{align}
& \sum_{g \in \mathcal{G}_i} P_{g,b,c} - \sum_{d \in \mathcal{D}_i} (P_{d,b} - P'_{d,b,c} ) = \sum_{ij \in \mathcal{L}_i} P_{ij,b,c} \pm P_{i,c}  
\nonumber \\
& \quad i \in \mathcal{V}, b \in \mathcal{B}, c \neq b_i \label{eq:PF-Pbalance}
\\
&\sum_{g \in \mathcal{G}_i} Q_{g,b,c} - \sum_{d \in \mathcal{D}_i} (Q_{d,b} - Q'_{d,b,c} ) = \sum_{ij \in \mathcal{L}_i} Q_{ij,b,c}  \pm
Q_{i,c} \nonumber \\ & 
\quad i \in \mathcal{V}, b \in \mathcal{B}, c \neq b_i 
\label{eq:PF-Qbalance} \\
&P_{ij,c}=g_{ii}V_{ij,c}+ g_{ij}\frac{V_{ij,c}-V_{ji,c} }{2} - b_{ij}(\theta_{ij,c}-\theta_{ji,c}) \nonumber \\& + P^L_{ij,c} \quad l_{ij} \in \mathcal{L},  c \neq l_{ij} \label{eq:Pij-AC} 
\end{align} 
\begin{align}
&Q_{ij,c}=-b_{ii}V_{ij,c}- b_{ij}\frac{V_{ij,c}-V_{ji,c} }{2} - g_{ij}(\theta_{ij,c}-\theta_{ji,c}) \nonumber \\
& + Q^L_{ij,c} \quad l_{ij} \in \mathcal{L},  c \neq l_{ij} \label{eq:Qij-AC} 
\\
& P^2_{ij,c}+Q^2_{ij,c} \leq \overline{S}^2_{ij} \quad l_{ij} \in \mathcal{L}  \label{eq:linelimit} \\
& \underline{V}_i  \leq V_{i,b,c} \leq \overline{V}_{i}  \quad i \in \mathcal{V},  b \in \mathcal{B} \label{eq:Vmax} 
\end{align}
\end{subequations}
where \eqref{eq:PF-Pbalance}-\eqref{eq:PF-Qbalance} present active and reactive power balance for each substation's busbars. Active and reactive power flows are defined in \eqref{eq:Pij-AC}-\eqref{eq:Qij-AC}, with $P^L_{ij,c},Q^L_{ij,c}$ representing active and reactive power losses.  \eqref{eq:linelimit} models the thermal limit of lines. \eqref{eq:Vmax} represents voltage limits. We use a piecewise linearization of quadratic constraint \eqref{eq:linelimit}, and a linear approximation of loss terms $P^L_{ij,c},Q^L_{ij,c}$ around an initial operating point as in \cite{yang2017linearized,alizadeh2022tractable}.

\subsubsection{Objective} 
The objective function \eqref{eq:cost} minimizes the total operational cost, including (re-)dispatch cost in normal state based on the electricity market values \cite{alizadeh2022tractable}, spinning reserve costs, and load shedding cost under contingency states, as shown below:
\begin{equation}    
     \label{eq:cost}
     \begin{split}         
     \min_{} \quad &  \sum_{g \in \mathcal{G}} \big(  \pi_g (P^0_{g} - \hat{P}^m_g)    + \pi^u_g r^u_g + \pi^d_g r^d_g \big)  \\
     &  + 
     \sum_{c \in \Omega^*} \sum_{b \in \mathcal{B}} \sum_{d \in \mathcal{D}} \pi_d P'_{d,b,c}  
     \\  &\text{s.t. \eqref{eqCO-sub}-\eqref{eqCO-PF} .}  
     \end{split}
\end{equation}
We refer to \eqref{eq:cost} as the security-constrained substation reconfiguration (SC-SR) problem, which considers N-1 line, coupler, and busbar contingencies, as well as linear AC PF equations in \eqref{eqCO-sub}-\eqref{eqCO-PF}.

\section{Multi-Master Heuristic Approach}
\label{sec3:solution}
The developed SC-SR problem is challenging to solve for large-scale systems due to the combinatorial explosion of binary variables and the high number of contingency states. This section proposes an efficient and scalable approach to address this challenge. First, we present a conventional Benders decomposition baseline, followed by our proposed heuristic decomposition approach with multiple master problems.

\subsection{Conventional Benders Decomposition}

Building on \cite{heidarifar2021optimal,khanabadi2012OTS}, we decompose the SC-SR problem into a master problem (MP) and sub-problems. The MP determines generator dispatch and substation topology under normal operation ($c=0$). Feasibility sub-problems (FSP) check N-1 line contingencies, while the optimality sub-problems (OSP) evaluate load shedding for coupler and busbar contingencies. We assume zero load shedding for N-1 non-radial line contingencies, though they can be included in the OSP if needed.

\subsubsection{Master problem}
The MP can be formulated as below:
\begin{equation}
\label{eq:BD-C-MP0}
\begin{split}    
    \min_{}  \quad \mathcal{O}^{MP} & = \sum_{g \in \mathcal{G}} \pi_g x_{g} + \sum_{c \in \Omega^*}\psi_c  \\
\text{s.t} \quad & \eqref{eqCO-sub}-\eqref{eqCO-PF} \quad c=0  \\
\end{split}
\end{equation}
where $\pi_g x_{g}$ is the first term in \eqref{eq:cost}, and $\psi_c$ is a positive auxiliary variables approximating OSPs. 
The solution of the MP, $\hat{x}=\{P^0_{g},r^u_g,r^d_g \}$ and $\hat{z}=\{ z_i, z_g, z_d, z_{ij} \}$ , is then fixed for sub-problems. The MP \eqref{eq:BD-C-MP0} is a relaxation of \eqref{eq:cost}, as it only considers the normal state and approximates the contingency states. Therefore, $\mathcal{O}^{MP}$ is a lower bound (LB) of the true objective in \eqref{eq:cost}. 

\subsubsection{Feasibility sub-problem}
Given a solution from the MP, the FSP checks its feasibility under N-1 line contingencies. The FSP is formulated by fixing the MP solution, introducing slack variables to \eqref{eq:PF-Pbalance}-\eqref{eq:PF-Qbalance}. The FSP objective $\mathcal{O}^{FSP}_c$ minimizes infeasibility by penalizing slack values for each line contingency $c \in \Omega^l$. In case of any infeasibility, a \textit{feasibility cut} is added to the MP.

\subsubsection{Optimality sub-problem}
Given a feasible solution from the MP, the OSP evaluates solution optimality for substation contingency states by:
\begin{subequations} 
    \begin{align}
\label{eq:OSP}
    \min_{} \quad & \mathcal{O}^{OSP}_c = \sum_{b \in \mathcal{B}}  \sum_{d \in \mathcal{D}} \pi_d P'_{d,b,c} \quad \quad c \in \Omega^*   \\
\text{s.t} \quad &  z_. = \hat{z}_. \leftrightarrow  \mu_{.,c} \label{eq:fix-zi} \\
 & x_. = \hat{x}_.  \leftrightarrow \lambda_{.,c}  \label{eq:fix-x} \\
& \text{and \eqref{eqCO-sub}-\eqref{eqCO-PF}} \nonumber
\end{align}
\end{subequations}
where \eqref{eq:OSP} minimizes load shedding for each contingency $c \in \Omega^*$. After solving the OSP, an upper bound (UB) on the objective can be defined to check the optimality gap of the solution:
\begin{subequations}
\label{eq:UBLB}
    \begin{align}
         LB = &\mathcal{O}^{MP} \\
         UB =& \sum_{g \in \mathcal{G}} \pi_g x_{g} + \sum_{c \in \Omega^*}\mathcal{O}^{OSP}_c
    \end{align}
\end{subequations}
If the gap $UB-LB\leq \epsilon^o$, then the algorithm stops. Otherwise, an \textit{optimality cut} is added to MP:
\begin{align} 
\label{original-OSP-cut}
     & \sum_{i \in \mathcal{V}}\mu_{i,c}(z_i - \hat{z}_i) + 
    \sum_{g \in \mathcal{G}}\mu_{g,c}(z_g - \hat{z}_g)  +
    \sum_{d \in \mathcal{D}}\mu_{d,c}(z_d - \hat{z}_d)  \nonumber  \\
    & + \sum_{ij \in \mathcal{L}}\mu_{ij,c}(z_{ij} - \hat{z}_{ij}) +
        \sum_{g \in \mathcal{G}} \Big( \lambda^p_{g,c}(P^0_{g} - \hat{P}^0_{g} )   \\
        & +  \lambda^{ru}_{g,c}(r^u_g - \hat{r}^u_g)+\lambda^{rd}_{g,c}(r^d_g - \hat{r}^d_g)   \Big)  \leq \psi_c  \quad c \in \Omega^*  \nonumber
\end{align}
$\lambda_{.,c},\mu_{.,c}$ refer to dual values for sub-problem c.

\subsubsection{Heuristic Cuts for Benders Decomposition}
The original cut in \eqref{original-OSP-cut} is not sufficiently tight, as the reconfiguration of substation $i$ has minimal impact on the load shedding of another substation contingency $c \in \Omega_j$. Thus, a tighter heuristic cut is proposed:
\begin{align}
\label{heuristic-OSP-cut}
     & \mu_{i,c}(z_i - \hat{z}_i) + 
    \sum_{g \in \mathcal{G}_i}\mu_{g,c}(z_g - \hat{z}_g)  +
    \sum_{d \in \mathcal{D}_i}\mu_{d,c}(z_d - \hat{z}_d)  \nonumber \\
    & + \sum_{ij \in \mathcal{L}_i}\mu_{ij,c}(z_{ij} - \hat{z}_{ij}) +
        \sum_{g \in \mathcal{G}}\Big( \lambda^p_{g,b,c}(P^0_{g} - \hat{P}^0_{g} )  \\
        &  \lambda^{ru}_{g,c}(r^u_g - \hat{r}^u_g)+\lambda^{rd}_{g,c}(r^d_g - \hat{r}^d_g)   \Big) \leq \psi_c  \quad i \in \mathcal{V}, c \in \Omega_i  \nonumber
\end{align}
where \eqref{heuristic-OSP-cut} only includes the topology terms of the substation where the contingency occurs. This results in a much tighter cut, improving Benders convergence; however, it leads to sub-optimality. 

\begin{figure}[t]
    \centering \includegraphics[width=0.45\textwidth, keepaspectratio=true,trim={0cm 0cm 17.5cm 0cm},clip]{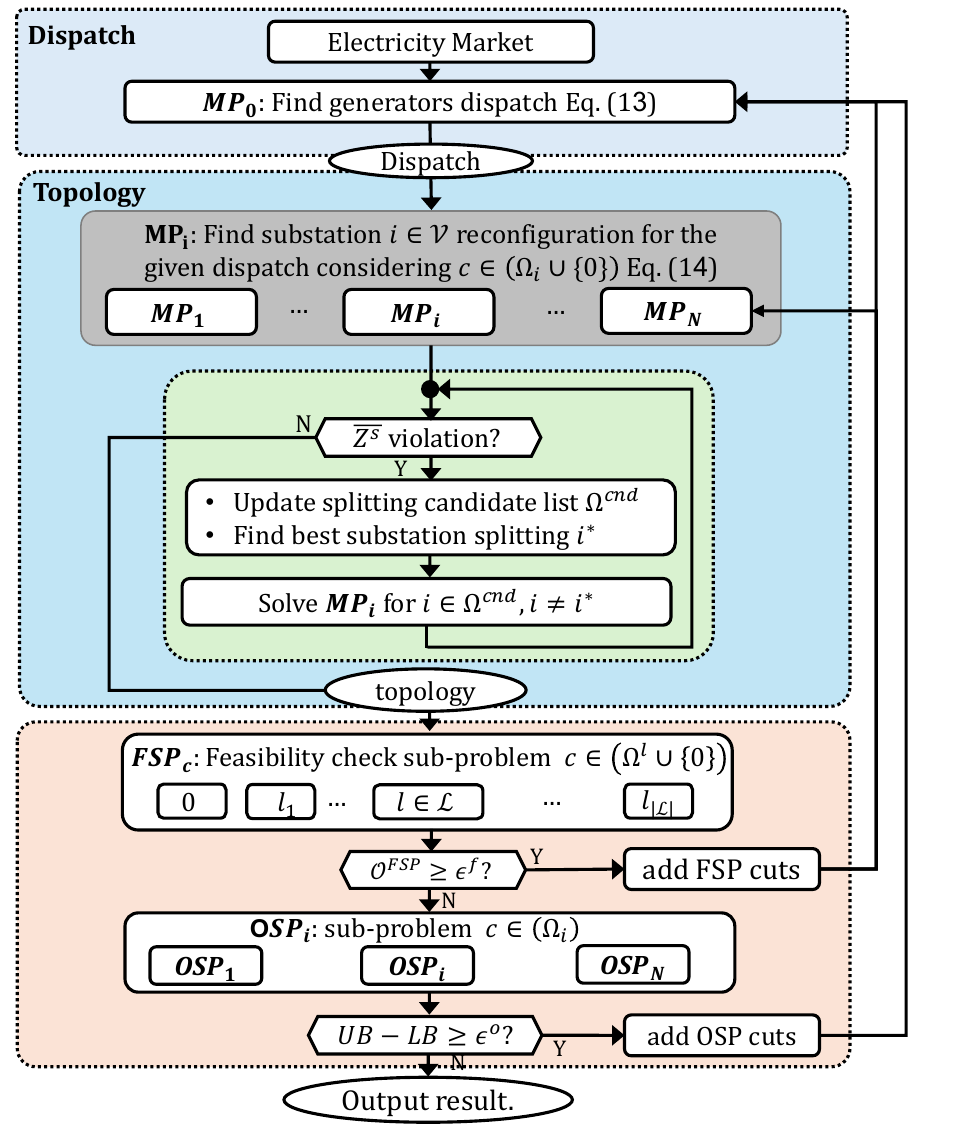}
    \caption{Flowchart of the proposed heuristic approach with multiple master problems.}
    \label{fig:flowchart}
\end{figure}

\subsection{The Proposed Heuristic Approach with Multiple Masters}
\cref{fig:flowchart} shows the proposed HMMP for solving SC-SR problem. HMMP combines concepts from BD, sequential heuristics, and problem partitioning in a unified decomposition tailored to SC-SR. The algorithm consists of two main MP stages: \textit{dispatch} and \textit{topology}. First, the optimal dispatch of generators are determined using an economic dispatch (ED) model $\mathit{MP}_0$. Second, multiple master problems $\mathit{MP}_i$ determine the topology of each substation in parallel. For preventive busbar splitting actions, a heuristic approach based on reducing objective value is proposed to sequentially identify optimal splitting actions. 
Finally, FSPs ensure feasibility for N-1 line contingencies, while OSPs evaluate load shedding for busbar and coupler contingencies, adding cuts if necessary.

Although HMMP is a heuristic that is not guaranteed to converge to the global optimum, it performs effectively for the SC-SR problem and provides computational speed-ups and practical applicability through the following: 
\begin{itemize}
    \item Decomposing the topology problem into independent substation problems $\mathit{MP}_i$, each with a small set of binaries, reduces complexity compared to conventional BD, which explores all binary combinations.
    \item $\mathit{MP}_i$ problems can be solved in parallel, improving scalability for large-scale grids.
    \item These speed-ups enable meeting market clearing time requirements. 
    \item Separating the dispatch and topology problems facilitates integration with operational and market practices.
\end{itemize}

The detailed steps of the proposed HMMP approach are outlined in Algorithm \ref{alg:proposed} and further explained:

\begin{algorithm}[t]
\caption{Proposed HMMP approach}
\label{alg:proposed}
\nonl \textbf{Dispatch}  \\
Receive initial dispatch $\hat{P}^m_g$ from electricity market \\
Solve $\mathit{MP}_0$ \eqref{eq:MP0} to determine dispatch $P^0_g,r^u_g,r^d_g$. \label{alg:MP0}\\
Fix dispatch as $\hat{x}$. \\
\nonl \textbf{Topology} \\
\For{$i \in \mathcal{V}$}{
Solve $\mathit{MP}_i$ \eqref{eq:MP_i} considering $c \in (\Omega_i \cup \{0\})$ to determine topology of substation $i$. 
}
\For{$sw \in \{0,...,\overline{Z}\}$ \label{alg:split0}}{
Update splitting candidate list $\Omega^{cnd}$. \\
\If{$ |\Omega^{cnd}|+|\Omega^{split}| > sw $ \label{alg:if_max_sw}}{
Add $i^*=\underset{i \in \Omega^{cnd}}{\mathrm{argmax}} \text{ } \Delta\mathcal{O}^{MP}_i $ to $\Omega^{split}$. \\
Fix $z_{i^*}=0$. \\
\For{$i \in \Omega^{cnd} , i\neq i^*$}{
Solve $\mathit{MP}_i$. \label{alg:split-end}
}
}
}
Fix topology as $\hat{z}$. \\
\For{$c \in (\Omega^l \cup \{0\})$ \label{alg:FSP-loop}}{
Solve $FSP_c$ and add feasibility cut to $\mathit{MP}_0$ (if $\mathcal{O}_c^{FSP} > \epsilon^f$). \label{alg:FSP-end} \\
}
\For{$i \in \mathcal{V}$}{
Solve $OSP_i$ \eqref{eq:OSP} considering $c \in \Omega_i$.
}
\eIf{$UB-LB > \epsilon^o$ \label{alg:ublb}}{
Add optimality cuts \eqref{OSP-cut-proposed} and go to line \ref{alg:MP0}. 
}{Output results.}
\end{algorithm}

\subsubsection{Dispatch Master problem $\mathit{MP}_0$}
At each iteration of the algorithm, $\mathit{MP}_0$ determines the dispatch by: 
\begin{subequations}
\label{eq:MP0}
    \begin{align}    
    \min_{} & \quad  \mathcal{O}^{MP}_0  =  \sum_{g \in \mathcal{G}} \pi_g x_g +   \sum_{i \in \mathcal{V}}\sum_{c \in \Omega_i}\psi_c  \end{align}
    \begin{align}
\text{s.t} \quad & \sum_{g \in \mathcal{G}}P^0_g \geq \sum_{d \in \mathcal{D}}\hat{P}_d \label{eq:MP0-ED} \\
& \underline{P}_g  \leq P^0_{g}-r^d_g, \quad P^0_{g}+r^u_g \leq \overline{P}_{g}  \quad g \in \mathcal{G}  \\
& \text{and \eqref{eq:g6}.} \nonumber
\end{align}
\end{subequations}
where $\eqref{eq:MP0-ED}$ ensures total supply-demand balance without considering grid constraints.

\subsubsection{Topology Master problem substation $i$ ($\mathit{MP}_i$)}
Given the dispatch solution from $\mathit{MP}_0$, each $\mathit{MP}_i$ independently determines the topology of substation $i$ in parallel:
\begin{subequations}
\label{eq:MP_i}
\begin{align}
    \min_{} & \quad \mathcal{O}^{MP}_i = \sum_{b \in \mathcal{B}}  \sum_{d \in \mathcal{D}} \sum_{c \in (\Omega_i \cup \{0\}) }  \pi_d P'_{d,b,c} \quad \quad  i \in \mathcal{V}   \\  
\text{s.t.} \quad & P^0_{g,b1} = (1-z_g)\hat{P}^0_g, \quad P^0_{g,b2} = z_g\hat{P}^0_g \quad g \in \mathcal{G}_i \label{eq:MPi-g0}
\\ & \text{\eqref{eqCO-sub}-\eqref{eqCO-PF} } \quad  c \in (\Omega_i \cup \{0\}) \nonumber
\end{align}
\end{subequations}
where \eqref{eq:MPi-g0} fixes the dispatch to identify the generator's connecting busbar. Notably, $\mathit{MP}_i$ considers only the contingencies associated with coupler and busbars of substation $i$ (i.e., $\Omega_i$) and the normal N-0 state. Thus, $\mathit{MP}_i$ identifies a secure and optimal substation reconfiguration that minimizes load shedding under these states ($\Omega_i \cup {0}$). Furthermore, while the FSP ensures feasibility under normal state, in case of a congestion, $\mathit{MP}_i$ addresses congestion by leveraging busbar splitting, enabling the realization of the optimal $\hat{P}_g^0$ dispatch and reducing the redispatch cost.

\begin{figure}[t]
    \centering \includegraphics[width=0.4\textwidth, keepaspectratio=true,trim={0cm 7cm 3cm 0cm},clip]{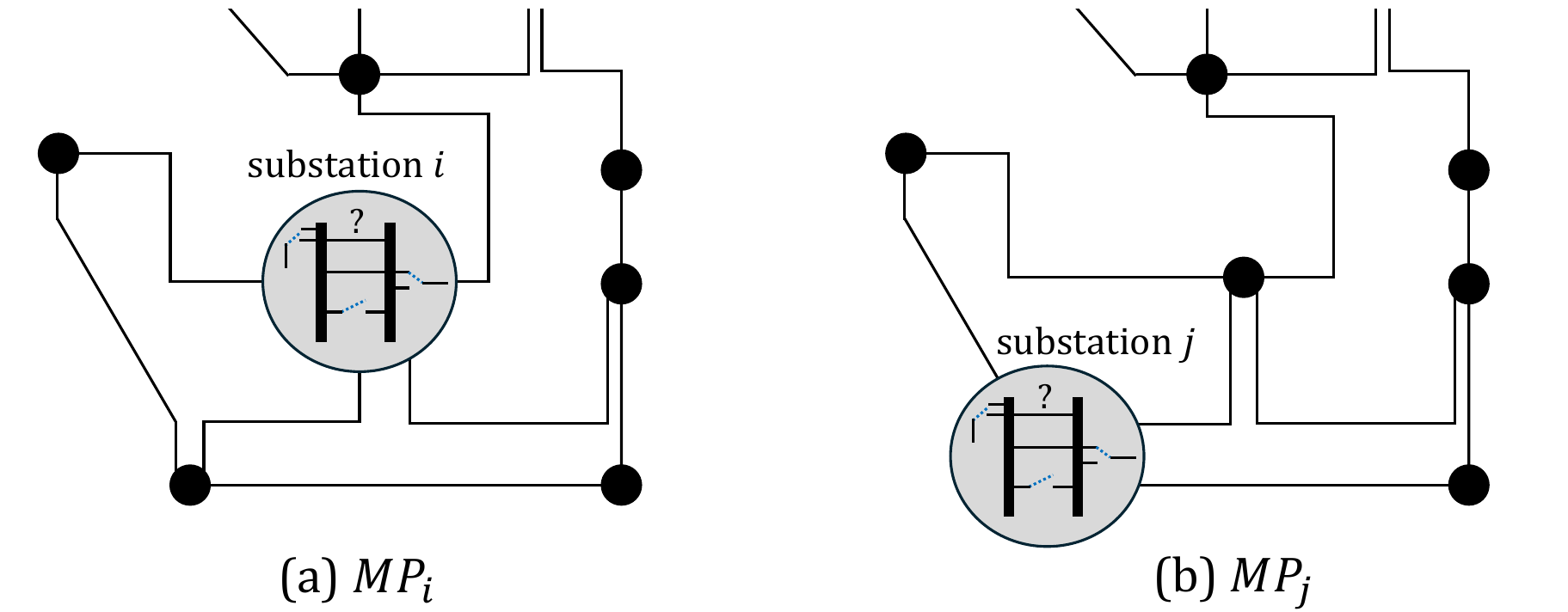}
    \caption{Schematic of the grid for each master problem: (a) MP for substation i, (b) MP for substation j.}
    \label{fig:Mpi_sub}
\end{figure}

As shown in \cref{fig:Mpi_sub}, $\mathit{MP}_i$ models all other substations $j \neq i$ as single electric nodes when they are not split, since in that case reconfiguring $j$ does not impact the objective $\mathcal{O}^{MP}_i$. If a substation $j$ has been split in a previous iteration, its configuration is fixed to the last determined topology while solving for $MP_i$. This modeling results in a significantly smaller MIP, and any resulting inaccuracy (for split substations) is often corrected through the iterative structure of HMMP. Additionally, feasibility is always checked using the full system configuration, and if a topology is infeasible, the corresponding FSP cut removes it from the solution space.

\subsubsection{Busbar splitting heuristic}
\label{sec:splitting}
As $\mathit{MP}_i$ problems determine each substation topology independently, the overall maximum splitting constraint \eqref{eq:zmax} may be violated. Additionally, if multiple splitting actions are selected, the system operator must determine their sequential order and ensure that none of these actions violate grid constraints. To this end, we propose a heuristic approach based on objective reduction to sequentially determine splitting actions one by one (lines \ref{alg:split0}-\ref{alg:split-end}). The heuristic begins with an empty set of splitting actions $\Omega^{split}$. Substations with $z_i=0, i \notin \Omega^{split}$ are added to the set of candidate substations $\Omega^{cnd}$ for splitting. Among the candidates, the substation that provides the maximum reduction in objective value is selected for splitting and added to $\Omega^{split}$:
\begin{subequations}
\label{eq:delta-obj}
\begin{align}
i^*= \underset{i \in \Omega^{cnd}}{\mathrm{argmax}} & \quad \Delta\mathcal{O}^{MP}_i \\
\textrm{s.t.} \quad & \Delta\mathcal{O}^{MP}_i = {\mathcal{O}^{MP}_i}^{(1)}-{\mathcal{O}^{MP}_i}^* \quad i \in \Omega^{cnd} 
\end{align}
\end{subequations}
where ${\mathcal{O}^{MP}_i}^*$ is the optimal objective of $\mathit{MP}_i$, and ${\mathcal{O}^{MP}_i}^{(1)}$ represents the initial objective obtained without busbar splitting (i.e., $z_i=0$) as returned by the MIP solver. The topology of the selected substation $i^*$ is then fixed, and the remaining candidates are solved again. This process iteratively identifies the next optimal splitting action until no further actions can be determined. However, as the splitting actions are selected greedily, the procedure is not guaranteed to identify the global optimal combination of actions, and different orders of splitting may lead to different solutions.

\subsubsection{Sub-Problems}

The FSPs check the feasibility of the dispatch and topology solution under normal and contingency conditions. If any constraint is violated, feasibility cuts are added to $\mathit{MP}_0$ (lines \ref{alg:FSP-loop}-\ref{alg:FSP-end}). Next, the OSP for all contingencies are solved to calculate the UB as in \eqref{eq:UBLB}. If the gap $UB-LB\leq \epsilon^o$, the algorithm terminates. Otherwise, the proposed heuristic optimality cut is added to $\mathit{MP}_0$ (line \ref{alg:ublb}):
\begin{align}
\label{OSP-cut-proposed}
    & \sum_{g \in \mathcal{G}}\big(\lambda^p_{g,c}(P^0_{g} - \hat{P}^0_{g} ) + \lambda^{ru}_{g,c}(r^u_{g} - \hat{r}^u_{g} )  \nonumber 
    \\  & + \lambda^{rd}_{g,c}(r^d_{g} - \hat{r}^d_{g} )\big)   \leq \psi_c   \quad i \in \mathcal{V}, c \in \Omega_i    
\end{align}

Regarding optimality guarantees, HMMP is a heuristic approach and is not guaranteed to converge to the global optimum. Separating the dispatch and topology problems may introduce sub-optimality. When no substation is split, dispatch and topology decisions are more \textit{decoupled} (e.g., changes in $z$ do not affect the bases-case PF), so the cut in \eqref{OSP-cut-proposed} is more effective; whereas splitting actions couple the two problems more strongly and may lead to higher optimality gaps. Nevertheless, this dispatch-topology decomposition aligns with the operational practice where dispatch and topology are determined in separate steps, and also fixing the dispatch enables each substation topology problem to be solved independently and in parallel, significantly reducing computational time.

\section{Case Studies}
\label{sec:results}
\subsection{Settings and Test Networks}
The case studies are performed on the IEEE 14-bus, 118-bus, and PEGASE 1354-bus transmission networks \cite{IEEEnetwork2019}. The substation model in \cref{fig:substation} is assumed for all nodes. The contingency list includes all transmission lines $\Omega^l$, couplers $\Omega^c$, and busbars $\Omega^b$. In Sec. \ref{sec:case-removeline}, 100 random samples are drawn from a Kumaraswamy(1.6,2.8) distribution with a Pearson correlation coefficient of 0.75 between loads \cite{giraud2024constraint}. In Sec. \ref{sec:case-dayahead}, the hourly load profile in \cite{saavedra2020day} is assumed. Big-M and voltage limit parameters are taken from \cite{heidarifar2021optimal}. For the linearization, the initial values of $\theta^0_{ij,c}$ and $v^0_{ij,c}$ are obtained by solving a lossless formulation of \eqref{eqCO-PF}, as in \cite{alizadeh2022tractable}. In all case studies, the solver is initialized with $\mathcal{T}_0$, where coupler is closed and all elements are connected to busbar 1. The average energy not supplied (ENS) is calculated as the average load shedding over contingencies divided by the total system load. Active contingencies refer to those in which any load shedding occurs.

The proposed formulations are implemented and solved with GurobiPy 9.5.2 \cite{gurobi} and Python 3.10. The MIPGap of Gurobi is set to 1\% and the time limit is 10 hours, if not specified. The numerical experiments are run on the DelftBlue supercomputer with Intel Xeon E5-6248R 24C 3.0GHz and 32GB RAM \cite{DHPC2022}. The following models are compared to investigate the efficacy of the proposed approach for SC-SR:
\begin{itemize}
    \item Org-MIP: the original MIP formulation \eqref{eq:cost} solved directly by Gurobi.
    \item BD-C: Conventional Benders decomposition.
    \item BD-H: Benders decomposition with heuristic cuts in \eqref{heuristic-OSP-cut}. 
    \item 1-Opt-H: an iterative one-step improvement heuristic that starts from the initial topology $\mathcal{T}_0$ and, at each iteration, flips a single binary variable that reduces the objective, until no improving move exists.
    \item Seq-H: a sequential heuristic that solves each substation topology problem ($MP_i$) one at a time while allowing the dispatch to update between iterations. Substation configurations are optimized in sequence until no further improvement is found.
    \item HMMP: the proposed approach in Algorithm \ref{alg:proposed}.
\end{itemize}

\subsection{Security with Substation Reconfiguration}
\label{sec:case-removeline}
This case study investigates how substation topology affects system security after a planned topology change. First, the optimal substation topology $\hat{\mathcal{T}}$ is determined by solving SC-SR \eqref{eq:cost} with all lines in service. Then, a random line is taken out of service as planned outage (e.g., for maintenance \cite{panel2021continental}). We compare: 1) the \textit{baseline} approach, representing current practice, which re-solves SC-OPF considering only N-1 line contingencies while keeping the substation topology $\hat{\mathcal{T}}$, 2) the proposed approach, which re-solves the SC-SR \eqref{eq:cost} to determine a new secure substation topology. Note that N-1 line contingencies represent unplanned outages, and are different from the planned outage of the removed line. No busbar splitting is considered here. 

\begin{figure}[t]
    \centering
    \includegraphics[width=0.4\textwidth]{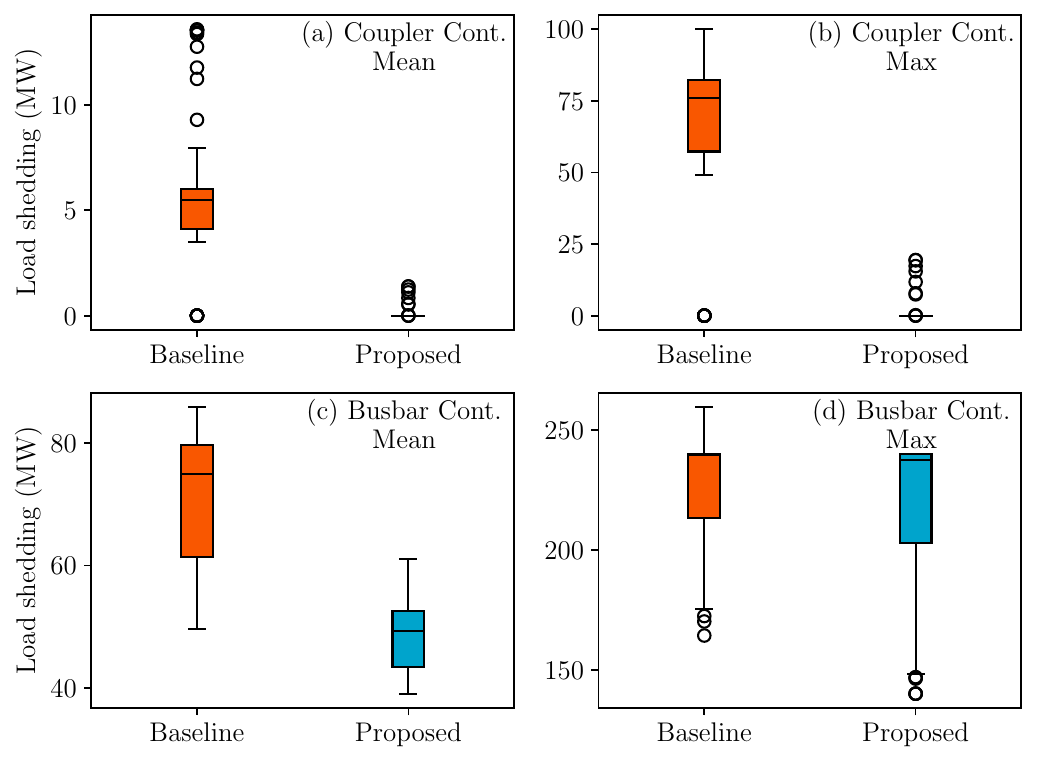}
    \caption{Mean and maximum load shedding over coupler and busbar contingencies for the baseline and proposed approach.}
    \label{fig:c1.1_line}
\end{figure}

\cref{fig:c1.1_line} presents the mean and maximum load shedding over coupler and busbar contingencies for the baseline and proposed approach, evaluated over 100 random load and line outage samples on the IEEE 14-bus system. \cref{fig:c1.1_line} (a)-(b) show that a single coupler outage can lead to significant load shedding, up to 25\% of system load, while re-optimizing the topology mitigates this risk. In \cref{fig:c1.1_line} (c), the proposed approach reduces the average load shedding for busbar contingencies by 50\% over the baseline. However, the maximum load shedding is not significantly improved with substation reconfiguration, due to significant loading on substation $i_2$, which can be addressed by reinforcing the substation with additional busbar sections. This case study highlights the need to optimize substation topology following a change in grid topology to ensure system security.

\subsection{Leveraging Parallel Computations}
\begin{figure}[t]
    \centering
    \includegraphics[width=0.36\textwidth]{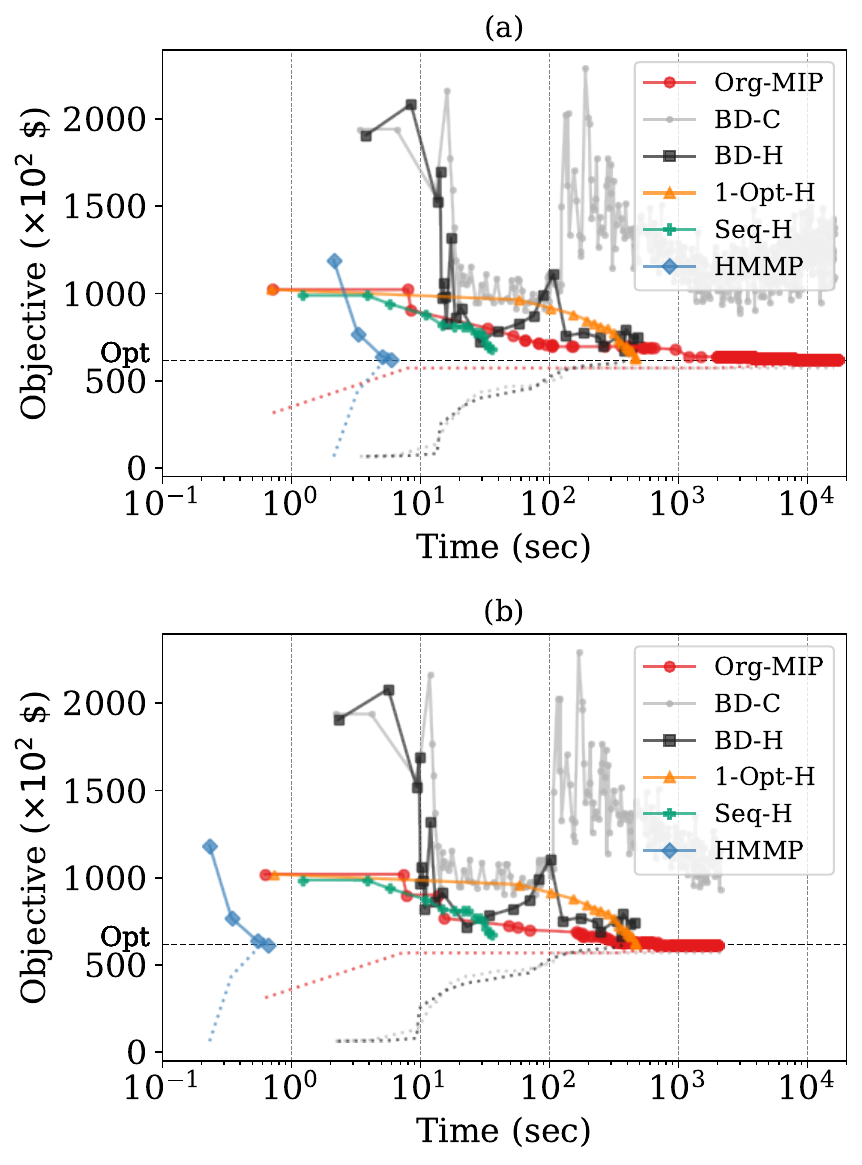}
    \caption{Objective convergence (UB and LB) on the 14-bus system: (a) Sequential computation with 1 CPU (32 GB RAM), (b) Parallel Computation with 16 CPUs (2 GB RAM each).}
    \label{fig:c2_ParSeq}
\end{figure}

This case study investigates the accuracy and computational efficiency of the proposed HMMP approach. \cref{fig:c2_ParSeq} presents the objective convergence (UB and LB) of the approaches over time on the 14-bus system with zero splitting. Computational time is analyzed for two scenarios: sequential (Seq.) and parallel (Par.) computations. In the Seq. case, a single CPU (32 GB RAM) is used, solving the SP of distributed approaches (BD-C, BD-H, and HMMP) sequentially. In the Par. case, 16 CPUs (2 GB RAM each) are used to solve SP of distributed approaches in parallel (in batches), while Gurobi is set to use 16 threads for Org-MIP. \cref{fig:c2_ParSeq} shows that Org-MIP achieves a 10\% optimality gap at a similar rate in both cases, but Par. reaches zero optimality gap about $10\times$ faster. For the BD models, the Par. computation of the SP does not reduce the time, as the main bottleneck remains the MP with binary variables. Moreover, 1-Opt-H and Seq-H show similar performance for Seq. and Par., as they cannot be parallelized. In contrast, the proposed HMMP is accelerated $10\times$, as the $\mathit{MP}_i$ problems are solved in parallel. Thus, we assume Par. computations for the rest of this paper.

Regarding convergence, BD-C fails to converge as loose cuts and symmetries in the problem prevent the MP from lowering the UB. BD-H with the heuristic cuts improves the convergence and outperforms Org-MIP in time, but results in some sub-optimality. 1-Opt-H reaches a close-to-optimal solution, but suffers from a low convergence rate as it evaluates all binary variables at each iteration to identify a single improving flip. Seq-H mitigates this by reconfiguring substations one at a time, supporting our hypothesis that decomposing the topology problem by substations is effective, but still leads to a sub-optimal solution. Finally, HMMP outperforms the baselines and achieves a high quality UB within a second.

\cref{table:14-bus} presents the optimality gap and computation time of approaches for different maximum splitting limits. With more splitting actions, the Org-MIP's computation time increases to 10 hours, as the solver explores more binary combinations. While BD-H computation remains relatively unchanged, its optimality gap increases up to 23\% due to heuristic cuts. Seq-H shows high optimality gaps, since the sequential order in which substations are processed can lead to different suboptimal solutions, particularly when splitting is allowed. HMMP and 1-Opt-H achieve notably lower gaps than the other baselines, with 2.5\% for the 2-split case. The higher gap for the 2-split case arises from the sequential, greedy nature of the splitting heuristic, which is not guaranteed to find the optimal combination of actions. Additionally, separating the dispatch and topology problems in HMMP can introduce larger gaps when more splitting actions are considered.

\begin{table}[t]
\renewcommand{\arraystretch}{1.2}
    \centering
\resizebox{0.48\textwidth}{!}{
\begin{tabular}{cc|cccccc}
\toprule
 & Max split & Org-MIP & BD-C & BD-H & 1-Opt-H & Seq-H & HMMP \\ \midrule
\multirow{3}{*}{\begin{tabular}[c]{@{}c@{}}Opt. \\ Gap \\ (\%)\end{tabular}} 
 & 0 split& - & 43.8 & 8.0 & 1.1 & 9.7 & 0.0 \\
 & 1 split & - & 43.2 & 23.2 & 2.1 & 8.7 & 0.9 \\
 & 2 split & - & 19.9 & 17.4 & 2.4 & 9.9 & 2.5 \\ \midrule
\multirow{3}{*}{Time} 
 & 0 split & 33 m & NA & 7.7 m & 7.7 m & 36 s & 0.5 s \\
 & 1 split & 5.7 h & NA & 6.0 m & 12.7 m & 41 s & 0.6 s \\
 & 2 split & 9.1 h & NA & 4.6 m & 11.0 m & 40 s & 0.6 s \\ \bottomrule
\end{tabular}
}
\caption{Performance comparison on the 14-bus system.}
\label{table:14-bus}
\end{table}

\subsection{Scalability to Large Systems}
\subsubsection{118-bus System}
\begin{figure}[t]
    \centering
    \includegraphics[width=0.36\textwidth]{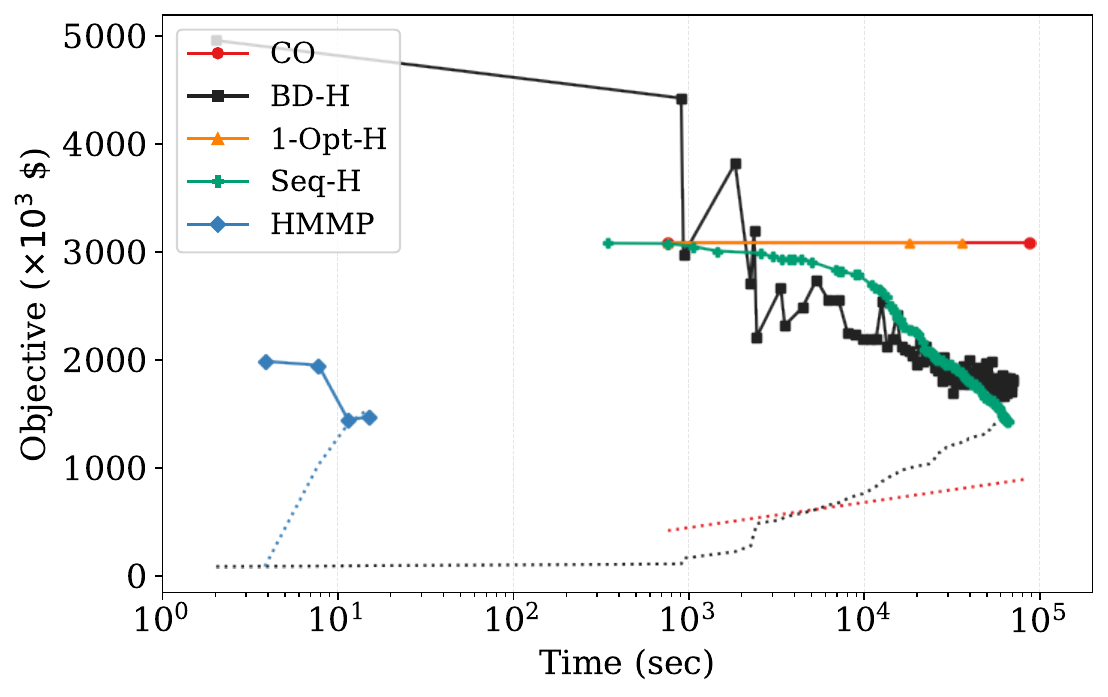}
    \caption{Objective convergence (UB and LB) on the 118-bus system.}
    \label{fig:c2_1118bus}
\end{figure}

\cref{fig:c2_1118bus} presents the objective (UB and LB) convergence performance of the approaches over time on the 118-bus system with zero splitting, while \cref{table:118} provides detailed performance metrics (e.g., number of curtailed loads). In \cref{fig:c2_1118bus}, Org-MIP fails to improve the UB from the initial topology $\mathcal{T}_0$ within 24 hours, due to the high number of binary variables and constraints. Similarly, 1-Opt-H does not converge as it fails to find the best optima at each iteration due to the high number of binaries. BD-H enhances convergence, improving the objective by 47\% and reducing the average ENS over contingencies from 11\% to 6\% of the total load. However, BD-H requires 100 iterations and approximately 20 hours. Seq-H improves the convergence rate, but still requires 18 hours as it cannot exploit parallel computation. In contrast, HMMP achieves objective improvement of 53\% while reducing computation time to under a minute by solving the MPs in parallel, providing fast and high-quality solutions. 

In \cref{table:118}, HMMP increases the number of active contingencies by reducing the stress in the most critical contingencies through better balancing the feeders between the busbars, which in turn reduces both the average load curtailment and the number of curtailed loads.

\begin{table}[t]
\renewcommand{\arraystretch}{1.1}
    \centering
\resizebox{0.48\textwidth}{!}{
\begin{tabular}{lccccc}
\toprule
 & Org-MIP & BD-H & 1-Opt-H & Seq-H & HMMP \\ \midrule
Obj. Improv. (\%) & - & 46.6 & 0.0 & 53.1 & 53.6 \\
Avg LS Cost (k\$) & 550.5 & 317.8 & 550.5 & 247.3 & 243.8 \\
\# Active Cont & 118 & 151 & 117 & 143 & 144 \\
Avg \#Curt. Loads & 2.5 & 1.4 & 3.7 & 1.7 & 1.7 \\
Avg LS (MW) & 55.0 & 31.8 & 55.0 & 24.7 & 24.4 \\
Avg LS$_{act}$ (MW)	& 251.9	& 113.7	& 254.1	 & 93.3	& 91.4 \\
Avg ENS (\%) & 1.3 & 0.7 & 1.3 & 0.6 & 0.6 \\
Time & $>$24 h & 21 h & $>$24 h & 18 h & 19 s \\ \bottomrule
\end{tabular}
}
\caption{Performance comparison on the 118-bus system.}
\label{table:118}
\end{table}

\subsubsection{1354-bus System}
\cref{table:1354bus-var} presents the computational complexity of the approaches for the 1354-bus system. While Org-MIP solves the problem as a single large MIP, BD considers the contingencies in the SP and reduces variables and constraints in the $MP$ by a factor of $5\times10^3$. However, the $MP$ remains computationally challenging due to the high number of binary variables. 1-Opt-H solves the same formulation as Org-MIP but updates one binary at a time, whereas Seq-H solves a sequence of substation problems. HMMP decomposes the $MP$ into multiple $\mathit{MP}_i$, which can be solved in parallel, each with only 4.6 binary variables on average. This significantly reduces complexity, as MIP complexity grows exponentially with binary variables. Additionally, the assumption in \cref{fig:Mpi_sub}, which models substations $j \neq i$ as single nodes, further reduces variables and constraints by a factor of 3, boosting efficiency.   

\cref{table:1354} compares Org-MIP and HMMP on the 1354-bus system. Notably, Org-MIP is intractable and encounters an out-of-memory error on a system with 1 TB RAM. An initial solution is obtained by fixing the topology to  $\mathcal{T}_0$, solving SC-OPF baseline considering line contingencies, and evaluating load sheddings for busbar contingencies. In contrast, HMMP effectively determines secure substation topologies, significantly reducing the objective by 65\% and lowering average load shedding by 330 MW, in 43 minutes. The MPs in HMMP are solved in batches of 16 (number of CPU cores), and more CPUs (e.g., 64 in modern servers \cite{DHPC2022}) can further speed up computations to a few minutes \cite{rajaei2025learning}. These results demonstrate the proposed approach's efficiency and scalability, essential for real-world operations and market compliance.

\begin{table}[t]
\renewcommand{\arraystretch}{1.0}
    \centering
\begin{tabular}{@{}cccccc@{}}
\toprule
\multicolumn{1}{l}{} &  & \multicolumn{2}{c}{No. of Variables} & \multirow{2}{*}{\begin{tabular}[c]{@{}c@{}}No. of \\ Constraints\end{tabular}} & \multirow{2}{*}{\begin{tabular}[c]{@{}c@{}}No. of\\ Opts.\end{tabular}} \\ \cmidrule(lr){3-4}
 &  & Binary & Continous &  &  \\ \midrule
Org-MIP & & 6217 & 2.8E+8 & 8.2E+8 & 1 \\ \midrule
\multirow{2}{*}{BD} & $MP$ & 6217 & 5.0E+4 & 1.4E+5 & 1 \\
 & $SP_c$ & - & 5.8E+4 & 1.3E+5 & 6053 \\ \midrule
1-Opt-H & & 6217 & 2.8E+8 & 8.2E+8 & 6217 \\ \midrule
Seq-H & & 4.6 & 7.5E+4 & 1.9E+5 & 1354  \\ \midrule
\multirow{3}{*}{HMMP} & $\mathit{MP}_0$ & - & 2.6E+3 & 1.0E+3 & 1 \\
 & $\mathit{MP}_i$ & 4.6 & 7.5E+4 & 1.9E+5 & 1354 \\
 & $SP_c$ & - & 1.8E+4 & 4.3E+4 & 6054 \\ \bottomrule
\end{tabular}
\caption{Number of variables and constraints for 1354-bus.}
\label{table:1354bus-var}
\end{table}

\begin{table}[t]
\renewcommand{\arraystretch}{1.1}
    \centering
\begin{tabular}{lcc}
\toprule
 & Org-MIP & HMMP \\ \midrule
Obj. Improv. (\%) & - & 60.6 \\
Avg LS Cost (k\$) & 2110.4 & 820.5 \\
\# Active Cont & 1354 & 2531 \\
Avg \#Curt. Loads & 3.2 & 21.4 \\
Avg LS (MW) & 211.0 & 82.1 \\
Avg LS$_{act}$ (MW) & 1077.1 & 196.2 \\
Avg ENS (\%) & 0.28 & 0.11 \\
Time & Out of Mem. & 38 m \\ \bottomrule
\end{tabular}
\caption{Performance comparison on the 1354-bus system.}
\label{table:1354}
\end{table}

\subsection{Application in Probabilistic Security}
This case study investigates the impact of a probabilistic approach on system security and cost. A probabilistic approach considers different contingency probabilities, minimizing the overall risk (probability$\times$consequence) \cite{ciapessoni2016probabilistic} and balancing cost and security. The objective in \eqref{eq:cost} is extended to consider a contingency probability $p_c$ in the second term (i.e., $p_c\pi_dP'_{d,b,c}$).

\begin{figure}[t]
    \centering
    \includegraphics[width=0.4\textwidth]{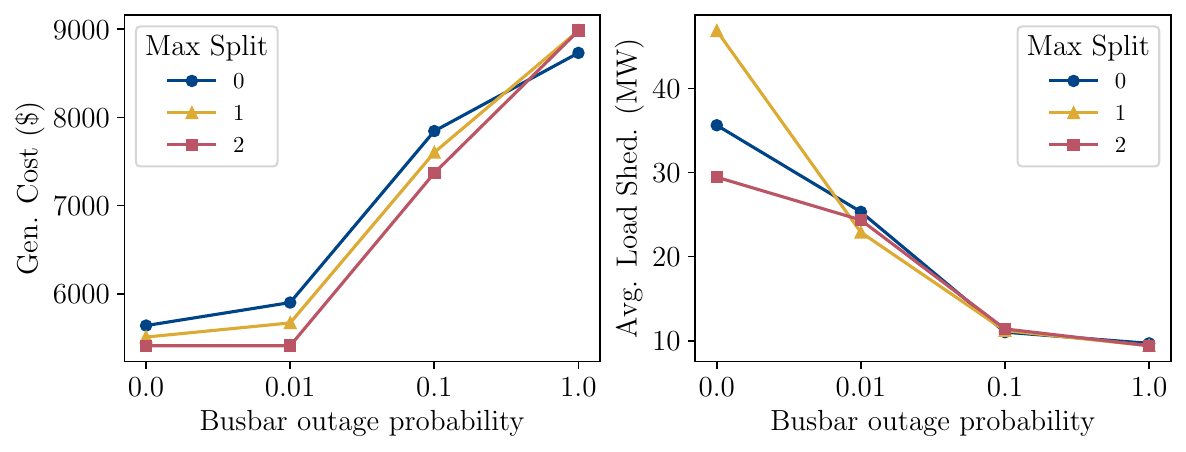}
    \caption{Generation cost and average load shedding over busbar outage probabilities on the 14-bus system.}
    \label{fig:c2.3_prob}
\end{figure}

\cref{fig:c2.3_prob} presents generation cost and average load shedding over busbar outage probabilities $p_b$ on the 14-bus system. We vary $p_b$ from 0.0 to 1.0 for the busbar contingencies, while coupler and line contingencies are assigned a probability of 1.0. Setting $p_b=0$ corresponds to an an optimal busbar splitting problem \cite{heidarifar2021optimal}. As $p_b$ increases, generation cost increases, and load shedding decreases. At $p_b=0.01$, load shedding drops by 30\%, with minimal cost increase. Busbar splitting reduces generation cost at lower probabilities, but may increase or not affect load shedding. At $p_b=1.0$, where load shedding dominates the objective, the solver does not select busbar splitting. These results indicate that busbar splitting effectively reduces cost and congestion, but offers limited benefit in reducing the impact of busbar outages.

\subsection{Application in Markets: Balancing Security and Cost}
In practice, markets determine generator dispatch, while operators ensure physical feasibility under normal and N-1 states. However, optimizing the system for busbar contingencies can significantly increase costs. To balance cost and security, we develop a \textit{fixed-cost} approach. We consider two cases: 
\begin{itemize}
    \item Fix the market dispatch ($P_g^0=\hat{P}_g^m$) and solve SC-SR,
    \item Impose an upper limit on the generation cost based on the market dispatch, and solve SC-SR. The objective in \eqref{eq:cost} is modified as: 
\begin{align}    
     \min_{} & \sum_{c \in \Omega^*} \sum_{b \in \mathcal{B}} \sum_{d \in \mathcal{D}} \pi_d P'_{d,b,c}  \\
     \text{s.t.} \quad  &\sum_{g \in \mathcal{G}} \big(  \pi_g P^0_{g}     + \pi^u_g r^u_g + \pi^d_g r^d_g \big) \leq  (1+\alpha)  \sum_{g \in \mathcal{G}} \pi_g P^m_{g}  \label{eq:gen-uppercost} \nonumber
\end{align}         
where $\alpha$ is a positive parameter representing the allowable increase in generation cost relative to market cost.
\end{itemize}

\begin{figure}[t]
    \centering
    \includegraphics[width=0.36\textwidth]{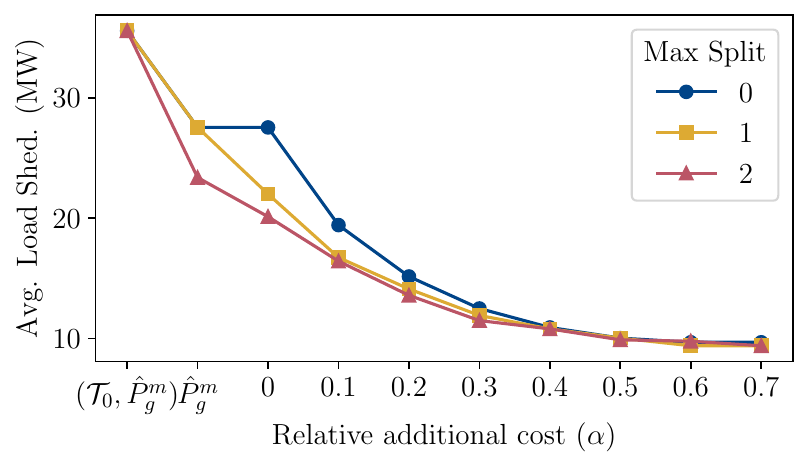}
    \caption{Load shedding vs operational cost on the 14-bus system.}
    \label{fig:c1.5_alpha}
\end{figure}

\cref{fig:c1.5_alpha} presents the average load shedding for different generation costs. To remove the impact of line contingencies, we solve an SC-OPF considering line contingencies as the market dispatch $\hat{P}_g^m$. Additionally, $\mathcal{T}_0$ represents the initial topology where all element are connected to busbar 1. The results show that substation reconfiguration reduces load shedding by 30\% for the fixed $\hat{P}_g^m$ dispatch compared to ($\mathcal{T}_0,\hat{P}_g^m$). Furthermore, busbar splitting further reduces load shedding by 20\% without increasing generation cost ($\alpha=0$), as it enables redistributing dispatch among generators with the same cost. This is in contrast to \cref{fig:c2.3_prob}, where busbar splitting, in some cases, increased load shedding. Therefore, operators can find the most secure topology without imposing additional operational cost to the system or requiring accurate risk modeling.

\subsection{Day-Ahead Scheduling of Substation Reconfiguration}
\label{sec:case-dayahead}

\begin{figure}[t]
    \centering
    \includegraphics[width=0.48\textwidth]{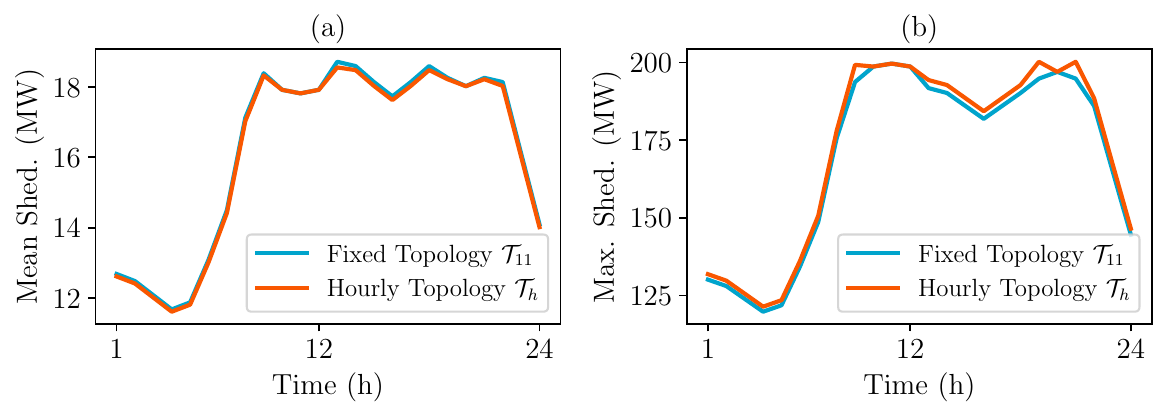}
    \caption{Load shedding over time for fixed and hourly topologies for the 14-bus system: (a) Mean values, (b) Maximum values.}
    \label{fig:c1.2_dayahead}
\end{figure}

This case study investigates the day-ahead scheduling of substation topology and its impact on load shedding. Two cases are considered: (1) Fixed topology, optimized for the peak hour and applied to the full day, and (2) Hourly topology, optimized per hour. \cref{fig:c1.2_dayahead} shows the mean and maximum load shedding over 24 hours. Loads are scaled using the hourly profile from \cite{saavedra2020day}, with peak demand at hour 11. Results show that hourly optimization provides minimal improvement over a fixed topology optimized for the critical hour. Combined with insights from \cref{fig:c1.1_line}, we conclude from our experiment that solving SC-SR hourly is unnecessary—secure topology can instead be determined day-ahead or after a topology change. 

SC-SR prioritizes system security by accounting for busbar outages, whereas OSR \cite{zhou2019bus,wang2023bus, van2023bus, heidarifar2015network,goldis2016shift, heidarifar2021optimal, morsy2022security,park2020optimal} focuses on minimizing operational costs. Our conclusion about the sufficiency of day-ahead scheduling applies to SC-SR. Hourly switching may still be beneficial in the context of OSR.

\subsection{Discussion}
The presented case studies demonstrate promising results for the proposed SC-SR formulation and HMMP solution approach. A key challenge for the baseline approaches is the symmetries inherent in SC-SR: reconfiguring substation $i$ has no or minor impact on the contingency states of substation $j$. This leads to stalled convergence in BD-C and 1-Opt-H, while BD-H and Seq-H partially mitigate these effects. HMMP overcomes this challenge by decomposing the topology problem across substations and solving the resulting $MP_i$ in parallel, providing substantial computational gains.

However, some limitations of HMMP can be noted. As a heuristic, it is not guaranteed to reach the global optimum. The greedy selection of splitting actions may miss beneficial combination of actions, which can lead to higher optimality gaps when multiple splits are allowed, as observed in \cref{table:14-bus}. Moreover, separating the dispatch and topology problem can facilitate integration with existing operational workflows and enables parallel computation, but this decomposition can lead to sub-optimality. In particular, when substations are split, the dispatch and topology problems become more \textit{coupled} (e.g., splitting affects the base-case PF), which can lead to higher optimality gaps. Furthermore, the AC PF equations \eqref{eqCO-PF} are linearized around an initial point; we refer to \cite{alizadeh2022tractable,yang2017linearized} for a discussion of how the initial point may influence the final solution.

This paper does not consider time-coupled resources, such as unit commitment or energy storage systems. HMMP can be extended to consider such resources and constraints. For a single day-ahead topology, all time-coupled variables and constraints can be included in the dispatch $MP_0$, the topology $MP_i$ can be solved for a representative hour (e.g., peak hour), and feasibility/optimality sub-problems are performed for all hours. If hourly topologies are allowed, $MP_i$ and the feasibility/optimality sub-problems are replicated for each hour of the horizon. 

HMMP provides a feasible solution at every iteration, and the objective cost improves with each iteration; operators can therefore stop the algorithm based on the UB–LB gap or their operational time requirements. Furthermore, implementing topology-optimization actions in real-world settings requires considering their impacts on protection and voltage stability schemes, while ensuring compliance with the structure, congestion management scheme, and operational rules of the target electricity market.

\section{Conclusion and Future Work}
\label{sec:conc}
This paper formulates the security-constrained substation reconfiguration (SC-SR) problem with linear AC PF equations, considering line, coupler, and busbar contingencies. Unlike existing methods that focus solely on optimal busbar splitting and neglect the topology of non-split substations, our approach determines a secure topology for all substations. Case studies demonstrate that SC-SR mitigates the impact of coupler tripping and reduces busbar-related load shedding by 50\% compared to an SC-OPF baseline. Moreover, results show that day-ahead substation reconfiguration is sufficient for maintaining system security, as hourly optimization provides minimal additional benefits in reducing load shedding. To balance system security and cost, we explore probabilistic and fixed-cost approaches.

Additionally, we propose a heuristic approach with multiple master problems (HMMP) to address the computational complexity of SC-SR. Case studies on the 118-bus and 1354-bus system show that HMMP achieves significant speed-ups, while maintaining minor optimality gaps compared to baselines included in this study (BD and iterative heuristic). The decomposition of MP to independent substation problems significantly reduces complexity, enables scalability to large systems, and allows for parallelization. This efficient approach paves the way for considering coupler and busbar contingencies into day-ahead operations, short-term maintenance and long-term expansion planning for large-scale power systems.

Future research can extend to more complex substation arrangements with multiple busbars, consider other contingencies (e.g. generators), and include time-coupled resources for real-world settings. Furthermore, the proposed  heuristic for identifying splitting actions and the dispatch-topology decomposition could be improved to reduce sub-optimality in cases requiring multiple splittings.


%


\appendix
\label{app-A}
As an example, consider a substation with four lines, $l_1,l_2,l_3,l_4$, connected to it. Without loss of generality, assume the busbars are split. The question is: how many unique ways the lines can connect to the busbars? \eqref{eq:z-sym0}-\eqref{eq:z-sym1} enforce that at least two lines must connect to each busbar. Let $(l_i,l_j)$ represent a configuration where $l_i$ connects to busbar $b1$ and $l_j$ connects to busbar $b2$. Therefore, the possible combinations are: $\{(l_1 l_2,l_3 l_4),(l_1 l_3,l_2 l_4),$ $(l_1 l_4,l_2 l_3),(l_2 l_3,l_1 l_4),(l_2 l_4,l_1 l_3), (l_3 l_4,l_1 l_2)\}$. However, note that there is a symmetry in the combinations; for instance, $(l_1 l_2,l_3 l_4)$ and $(l_3 l_4,l_1 l_2)$ are effectively identical. To this end, \eqref{eq:z-sym2} eliminates this symmetry by fixing one line (here the line with the lowest index) to $b1$. As a result, the following unique combinations remain feasible: $\{(l_1 l_2,l_3 l_4),(l_1 l_3,l_2 l_4),(l_1 l_4,l_2 l_3)\}$. Therefore, \eqref{eq:z-sym2} reduces the number of possible combinations by half.

\section*{Acknowledgment}
The authors are thankful to Jan Viebahn from TenneT, Antoine Marot,
and Benjamin Donnot from Reseau de Transport d’ Electricite, Benoît Jeanson from CRESYM, and Basel Morsy from Austrian Institute of Technology who provided valuable discussions.

\ifCLASSOPTIONcaptionsoff
  \newpage
\fi

\newcommand{\BIBdecl}{\setlength{\itemsep}{-0.0em}}

\bibliographystyle{IEEEtran}
\bibliography{Library}

\begin{IEEEbiography}[{\includegraphics[width=1in,height=1.25in,clip,keepaspectratio]{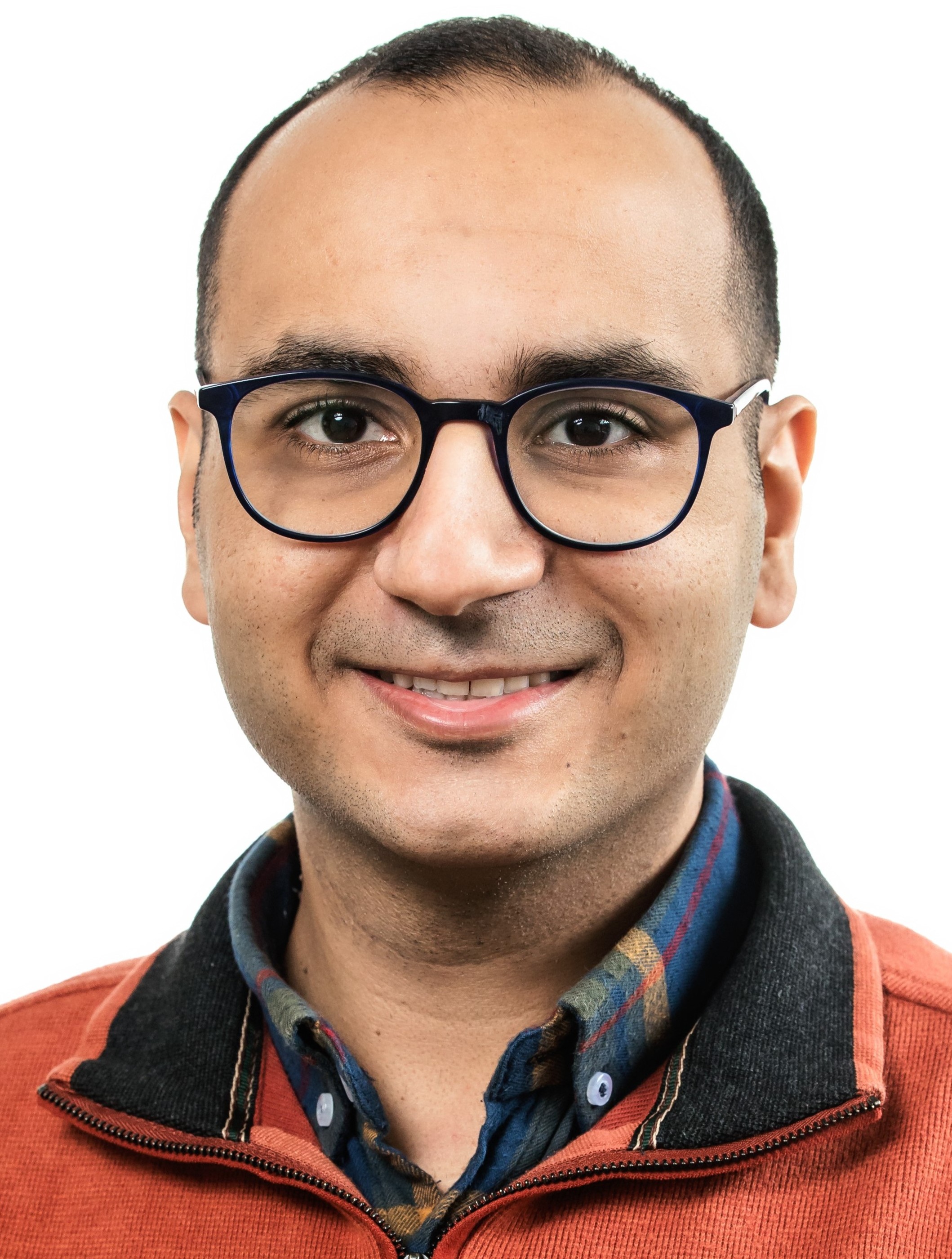}}]{Ali Rajaei}
(Graduate Student Member, IEEE) received the B.Sc. and M.Sc. degrees in electrical engineering from Sharif University of Technology, Tehran, Iran, in 2017 and 2019, respectively. He is currently a Ph.D. candidate with the Delft AI Energy Lab, Faculty of Electrical Engineering, Mathematics, and Computer Science, Delft University of Technology, Delft, The Netherlands. His research interests include machine learning and optimization for power system operation.
\end{IEEEbiography}

\begin{IEEEbiography}[{\includegraphics[width=1in,height=1.25in,clip,keepaspectratio]{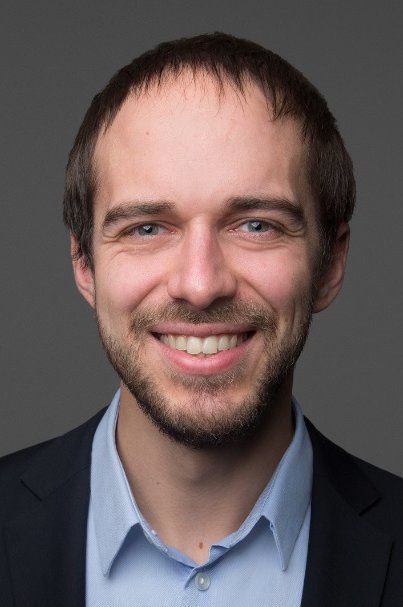}}]{Jochen Lorenz Cremer}
(Member, IEEE) received the Ph.D. degree from Imperial College London, U.K., in 2020. He works as an Associate Professor with the Delft AI Energy Lab, Faculty of Electrical Engineering, Mathematics, and Computer Science, Delft University of Technology, The Netherlands. His research interests include machine learning and mathematical programming applied to the operation and planning of power systems.
\end{IEEEbiography}

\end{document}